\documentclass[useAMS,usenatbib]{mn2e}

\usepackage[dvips]{graphicx}   

\title[The anatomy of a quadruply imaged gravitational lens system]{The anatomy of a quadruply imaged gravitational lens system}
\author[S. H. Suyu and R. D. Blandford]{S. H. Suyu$^{1,2}$\thanks{E-mail:suyu@its.caltech.edu} and R. D. Blandford$^{1,2}$\thanks{E-mail:rdb3@stanford.edu} \\
$^{1}$Theoretical Astrophysics, 103-33, California Institute of Technology, Pasadena, CA, 91125, USA \\
$^{2}$KIPAC, Stanford University, 2575 Sand Hill Road, Menlo Park, CA 94025, USA}

\begin{document}

\newcommand{\bd}{\begin{displaymath}}
\newcommand{\ed}{\end{displaymath}}
\newcommand{\be}{\begin{equation}}
\newcommand{\ee}{\end{equation}}
\newcommand{\beaa}{\begin{eqnarray*}}
\newcommand{\eeaa}{\end{eqnarray*}}
\newcommand{\bea}{\begin{eqnarray}}
\newcommand{\eea}{\end{eqnarrray}}

\date{20 October 2005}

\pagerange{000--000} \pubyear{2005}

\maketitle

\label{firstpage}

\begin{abstract}
The key to using a strong gravitational lens system to measure the Hubble constant is to obtain an accurate model of the lens potential.  In this paper, we investigate the properties of gravitational lens B1608+656, a quadruply-imaged lens system with an extended source intensity distribution.  Our analysis is valid for generic quadruply-lensed systems.  Limit curves and isophotal separatrices are defined for such systems, and we show that the isophotal separatrices must intersect at the critical curves and the satellite isophotes must be tangent to the limit curves.  The current model of B1608+656 \citep{K03} satisfies these criteria for some, but not all, of the isophotal separatrices within the observational uncertainty.  We study a non-parametric method of potential reconstruction proposed by \citet*{BSK01} and demonstrate that although the method works in principle and elucidates image formation, the initial potential only converges to the true model when it is within $\sim 1$ percent of the true model. 
\end{abstract}


\section{Introduction}
Strong gravitational lens systems provide a tool for measuring cosmological parameters.  With the measured relative arrival time delays between the multiple images of the lensed source and a model of the lens potential, one can deduce a value of the Hubble constant.  In addition, strong gravitational lens systems can be used to probe galaxy mass distributions, including dark matter, since the lens potential is directly related to the lens mass distribution. \citep{R64}

Several strong gravitational lenses with either two images (``doubles'') or four images (``quads'') have been observed.  The ones with extended source distributions are of special interest since they provide additional constraints for the lens potential due to surface brightness conservation.  The traditional approach to modelling the lens mass distribution is to postulate a parametric form for the lens distribution and minimize some chi-square to fit the data.  The method is limited by the choice of the parameters; as the observational quality improves, the original parametric model generally becomes inadequate to fit the data (\citet{WS00} consider a pixellated mass distribution which is non-parametric, but use only the nuclear image positions and not information from the extended source to constrain the distribution).  Ideally, we want a method that employs the extended source information to obtain a non-parametric form of the lens potential whose accuracy is limited solely by the observational noise in the data.  \cite{K05} has also taken this approach.

In section 2, we study in detail a quadruply imaged gravitational lens system with an extended source, B1608+656, showing the various criteria that the isophotes of the extended source must satisfy.  In section 3, we examine a method of potential reconstruction proposed by \citet{BSK01} to correct the potential values pixel by pixel from a starting perturbed potential model.

\section[]{Properties of a Quadruply Lensed System}

We will focus on the quadruply imaged gravitational lens system B1608+656 in this section.  Fig.~\ref{fig:HSTimage} shows an image of the system taken by the \textit{Hubble Space Telescope} through the F814W filter \citep{SB03}.  The source is at a redshift of $z_s = 1.39$ \citep{F96} and its images are labelled by A, B, C, and D.  The system has two lens galaxies, G1 (the primary lens) and G2 (the secondary lens), that are at a redshift of $z_d = 0.63$ \citep{M95}.\footnote{The quoted redshift is that of G1.  We assume that G2 is at the same redshift as G1 since G2 is too faint for its redshift to be measured.}  The galaxy G1 is about five times more massive than G2.  B1608+656 is unique in that all three relative time delays between the four images are determined with accuracies of a few percent.  The time delays relative to image B for images A, C, and D are $31.5 \pm 1.5$ days, $36.0 \pm 1.5$ days, and $77.0 \pm 1.5$ days, respectively \citep{F99,F02}.

\begin{figure}
\vspace{0.in}
\includegraphics[width=84mm]{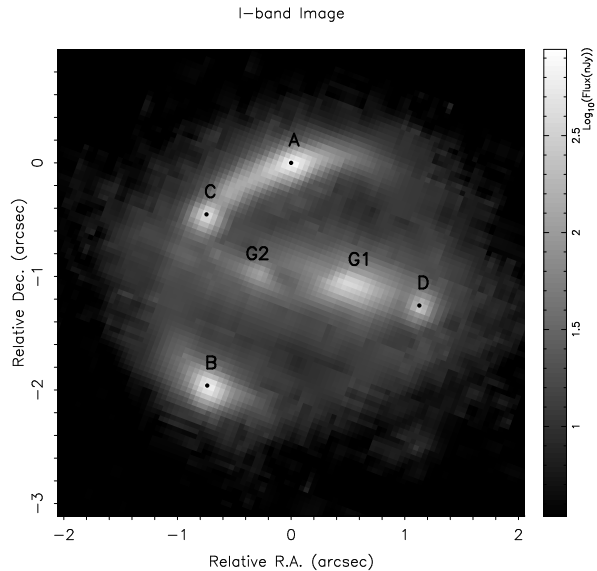}
\caption{\label{fig:HSTimage} The original reduced \textit{HST}/F814W image of B1608+656.  The four images are labelled A, B, C, and D; the two lens galaxies are G1 and G2.  (\citealt{SB03}; similar image in \citealt{K03})}   
\end{figure}
 
Section 2.1 that follows is a review of the theory of gravitational lensing.  Readers familiar with lensing may wish to proceed directly to Section 2.2, which analyses the \citet{K03} model of B1608+656.

\subsection{Gravitational lensing}

Readers familiar with gravitational lensing may wish to skip this section.  We follow \citet*{KSW} for the theory of gravitational lensing.  

Let us denote the angular coordinates on the source and image planes by $\bmath{\beta}=(\beta_1,\beta_2)$ and $\bmath{\theta}=(\theta_1,\theta_2)$, respectively.  The lens equation governing the deflection of light rays is
\be 
\label{eq:lensEq} 
\bmath{\beta} = \bmath{\theta} - \balpha(\bmath{\theta}),  
\ee
where $\bmath{\alpha}(\bmath{\theta})$ is the scaled deflection angle that is the gradient of a scalar function called the lens (or deflection) potential: 
\be
\label{eq:alpha}
\bmath{\alpha}(\bmath{\theta}) = \bmath{\nabla} \psi(\bmath{\theta}).
\ee
In terms of the dimensionless surface mass density, denoted by $\kappa(\bmath{\theta})$, the lens potential is 
\be \label{eq:psi} 
\psi(\bmath{\theta}) = \frac{1}{\pi} \int_{\Re^2} d^2\theta' \kappa(\bmath{\theta'}) \ln |\bmath{\theta}-\bmath{\theta'}|. 
\ee

The time delay function relative to the case of no lensing is 
\be \label{eq:T} 
T(\bmath{\theta},\bmath{\beta}) = \frac {1}{c} \frac{D_d D_s}{D_{ds}} (1+z_d) \left[\frac{(\bmath{\theta}-\bmath{\beta})^2}{2}-\psi(\bmath{\theta}) \right], 
\ee
where $D_{d}$, $D_{s}$, and $D_{ds}$ are, respectively, the angular diameter distance from us to the lens, from us to the source, and from the lens to the source.

The constant coefficient in equation (\ref{eq:T}) is proportional to the angular diameter distance and hence inversely proportional to the Hubble constant in a flat $\Lambda$-CDM universe.  Therefore, by measuring the relative time delays between the various images, we can in principle deduce the value of the Hubble constant if we know the source position ($\bmath{\beta}$) and the lens potential ($\psi(\bmath{\theta})$).  

To characterise the magnifications of images in gravitational lensing, a Hessian is used
\be \label{eq:jacob} 
\mathbfss{A}(\bmath{\theta}) = \frac {\partial \bmath{\beta}} {\partial \bmath{\theta}}. 
\ee
Using the lens equation (\ref{eq:lensEq}), the above equation can be written as
\begin{eqnarray} \label{eq:jacobMat}
\mathbfss{A}(\bmath{\theta}) & = & \left( \begin{array}{cc} 1-\psi_{11}(\bmath{\theta}) & -\psi_{12}(\bmath{\theta}) \\ -\psi_{12}(\bmath{\theta}) & 1-\psi_{22}(\bmath{\theta}) \end{array} \right), 
\end{eqnarray}
where the subscript 1 (or 2) in $\psi$ indicates a derivative with respect to $\theta_1$ (or $\theta_2$).
The magnification matrix is defined as $\bmath{\mu} = \mathbfss{A}^{-1}$, and the associated magnification factor is 
\be \label{eq:mag}
\mu(\bmath{\theta}) = \frac{1}{\det \mathbfss{A}(\bmath{\theta})}. 
\ee
According to equation (\ref{eq:mag}), the positions $\bmath{\theta}$ with $\det \mathbfss{A}(\bmath{\theta}) = 0$ have divergent magnification; the loci of such points on the image plane define the \textit{critical curves}.  Using the lens equation (\ref{eq:lensEq}), critical curves on the image plane are mapped to \textit{caustic curves} (or simply \textit{caustics}) on the source plane.  The caustic curves separate regions of different image multiplicities.

\subsection{Gravitational lens B1608+656}

To investigate the anatomy of the quad B1608+656, we use the mass distribution model proposed by \citet{K03}.  The parametric form of the dimensionless surface mass density for each of the two lens galaxies is a singular power law ellipsoid:
\be \label{eq:K03kappa} 
\kappa(\theta_{gal_1},\theta_{gal_2}) = b \left[ \theta_{gal_1}^2 + \left(\frac{\theta_{gal_2}}{q_l} \right)^2 \right]^{\frac{1-\gamma'}{2}},  
\ee
where $(\theta_{gal_1},\theta_{gal_2})$ are coordinates relative to the galaxy centre and $b$, $q_l$, and $\gamma'$ are parameters to fit the data.  The origin of coordinate $\bmath{\theta}$ is set at the position of image A.  Each of the lens galaxies is centred at the coordinates $(\theta_{l1},\theta_{l2})$ and is rotated by a major-axis position angle $\theta_{PA}$ that is measured from north to east (top to left).  There is an additional external shear centred on G1 whose contribution to the lensing potential, in polar coordinates relative to the shear centre ($(r,\phi)$ with $\theta_{sh_1}= r \cos(\phi)$ and $\theta_{sh_2} = r \sin(\phi)$), is
\be \label{eq:K03shear} 
\psi_{ext}(\bmath{\theta_{sh}}) = \frac{1}{2} \gamma_{ext} r^2 \cos(2 \phi), 
\ee
where $\gamma_{ext}$ is a parameter characterising the shear strength.  The rotation of the external shear is given by the position angle $\theta_{ext}$.  We adopt the parameter values of the SPLE1+D(isotropic) model in \citet{K03} and list them in Table \ref{tab:SPLE1}.

\begin{table}
\caption{\label{tab:SPLE1} Values for parameters in equations (\ref{eq:K03kappa}) and (\ref{eq:K03shear}) for the B1608+656 SPLE1+D(isotropic) model in \citet{K03}}
\begin{tabular}{|c|c|c|}
\hline
lens galaxy  & G1 & G2 \\
\hline 
$b$ & 0.526 & 0.269 \\
$q_l$ & 0.604 & 0.318 \\
$\gamma'$ & 2.05 & 2.12 \\
centroid $(\theta_{l1},\theta_{l2})$ & (0.425, -1.069) & (-0.291, -0.928) \\
position angle $\theta_{PA}$ ($^\circ$) & 77.0 & 68.4 \\
\hline
$\gamma_{ext}$ & \multicolumn{2}{c|}{0.077} \\
shear position angle $\theta_{ext} (^\circ)$  & \multicolumn{2}{c|}{13.4} \\
\hline 
\end{tabular}
\end{table}

\subsubsection{Critical and caustic curves}

The critical curves on the image plane and the caustic curves on the source plane of the SPLE1+D(isotropic) model in \citet{K03} are shown in Fig.~\ref{fig:imageTD2} in the middle panel and the left panel, respectively.  The locations of the lens galaxies are indicated by open triangles on the image plane.  The marked source and image locations will be discussed in the next section.  With the two elliptical lens galaxies, the large critical curve loop is a deformed version of an elliptical curve of one singular power law ellipsoid (equation (\ref{eq:K03kappa})).  The corresponding diamond shaped caustic curve, known as an astroid, is typical for elliptical mass distributions.  An astroid is composed of four folds (branches of smooth curves) joining at four cusps.  An individual power law ellipsoid has an astroid that is symmetrical with respect to the semi-major and semi-minor axis of the lens.  With the two lens galaxies in the SPLE1+D(isotropic) model, we have an asymmetry in the astroid and two additional small triangular caustics, called the deltoids, that map into the small loops on the image plane.

\subsubsection{Image positions and time delay surface}

It is instructive to see how the images move on the image plane as the source is displaced.  Understanding such movements is important for analysing quads and for defining the limit curves in the next section.  Fig.~\ref{fig:imageTD2} shows the locations of the images, labelled by A, B, C, D, and E (middle panel), when the source is at the centre of the astroid caustic (left panel).  Despite having five images, the system is called a quad because the central image is usually de-magnified and lies near the lens galaxies, making it nearly observationally invisible\footnote{We refer the reader to \citet*{WRK04} for candidates of central image detections in gravitational lens systems.}.  The arrival time delay contours in the right panel show that the image locations are at the time delay extrema or saddles, except for the extrema where the surface mass densities are non-smooth \citep{KSW}.  At the centroids of G1 and G2 whose locations are given in Table \ref{tab:SPLE1}, the time delay achieves local maxima, but there are no corresponding images because the surface mass densities are singular at the galaxy centroids in the model described by equation (\ref{eq:K03kappa}).  Ignoring the central image (E, which is finitely de-magnified), the two images (C and D) inside the critical curve are time delay saddles, and the two images (A and B) outside the critical curve are time delay minima.  This is true in general for quads.

\begin{figure*}
\includegraphics[width=56mm]{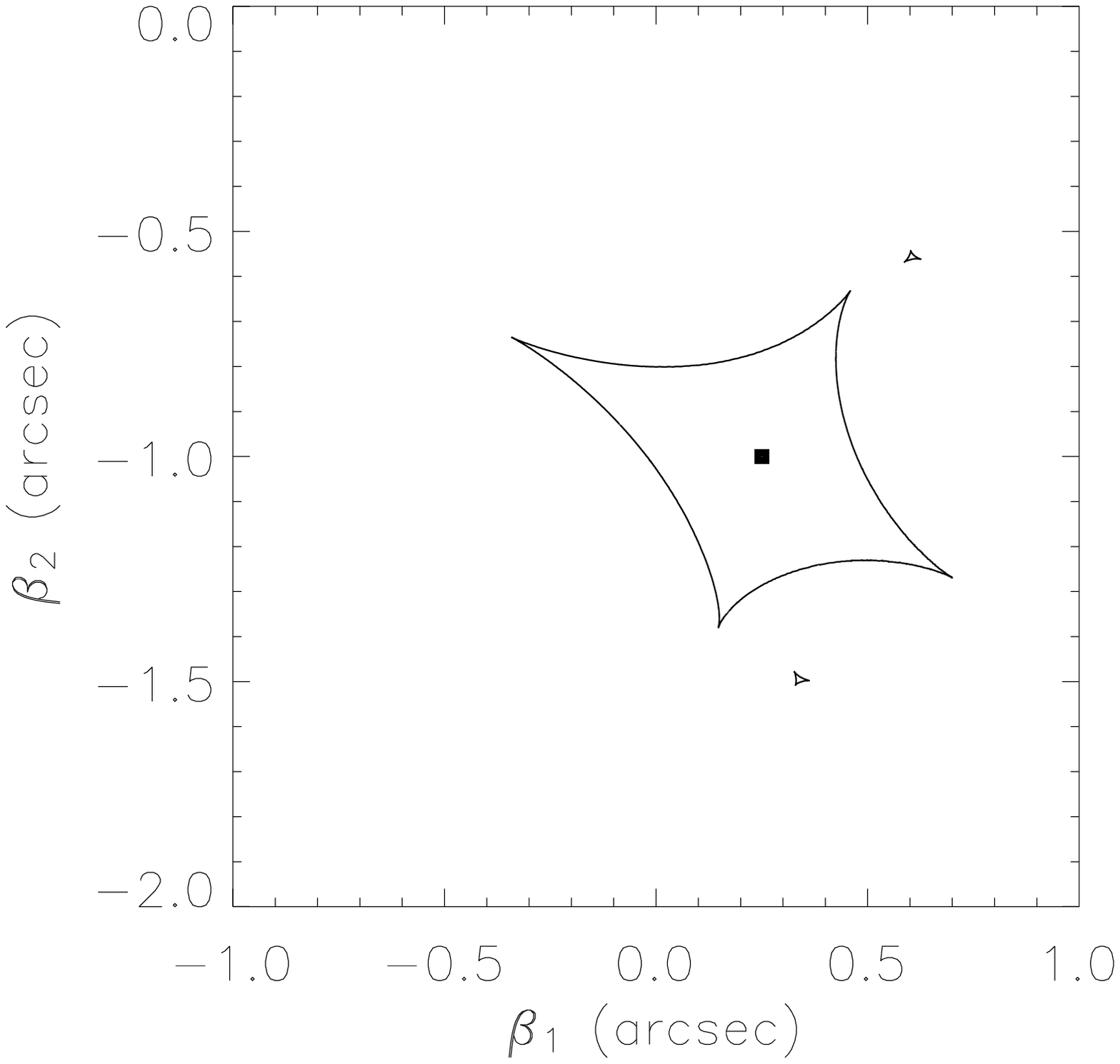}
\hspace{0.1in}
\includegraphics[width=56mm]{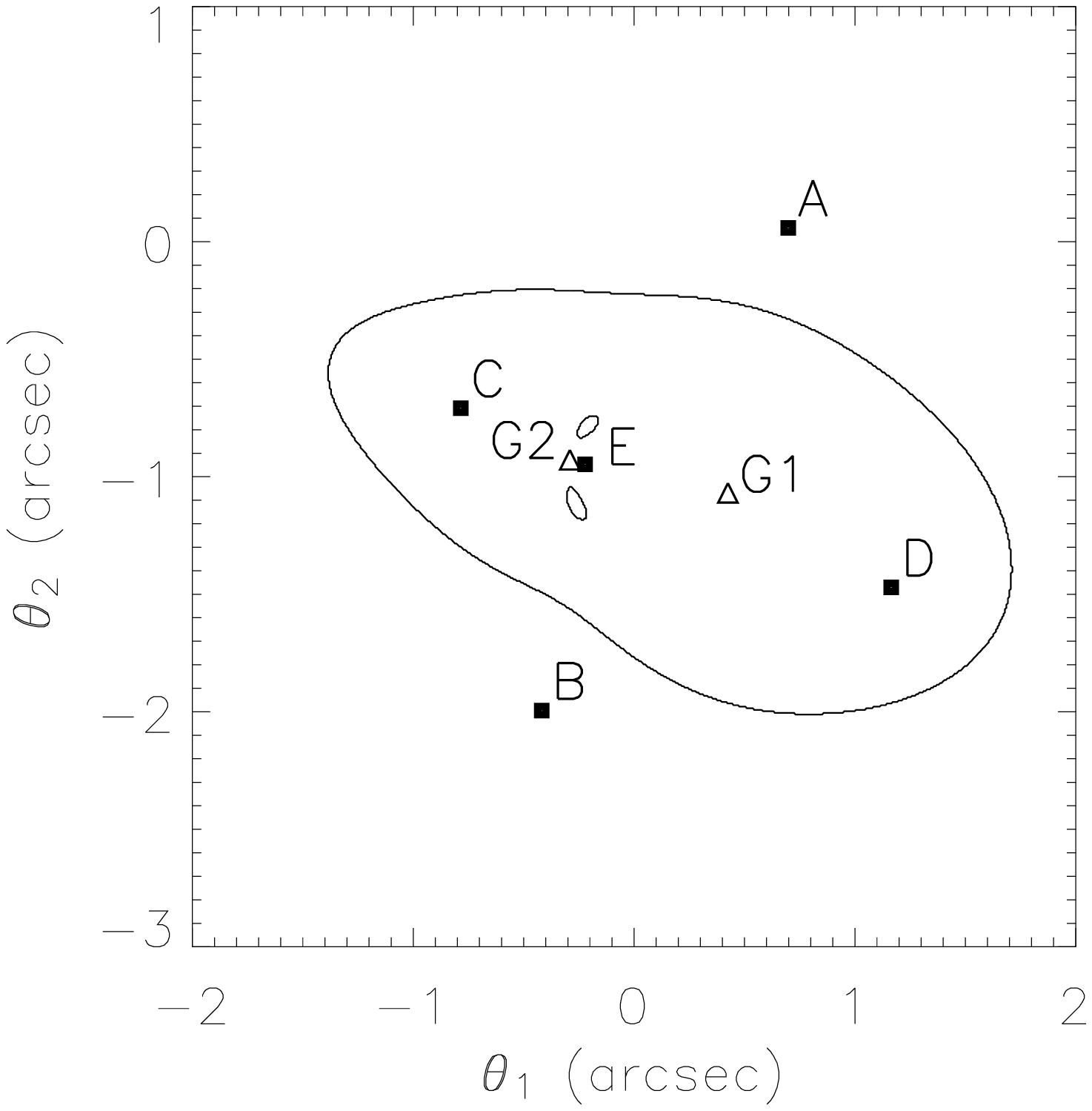}
\hspace{0.1in}
\includegraphics[width=56mm]{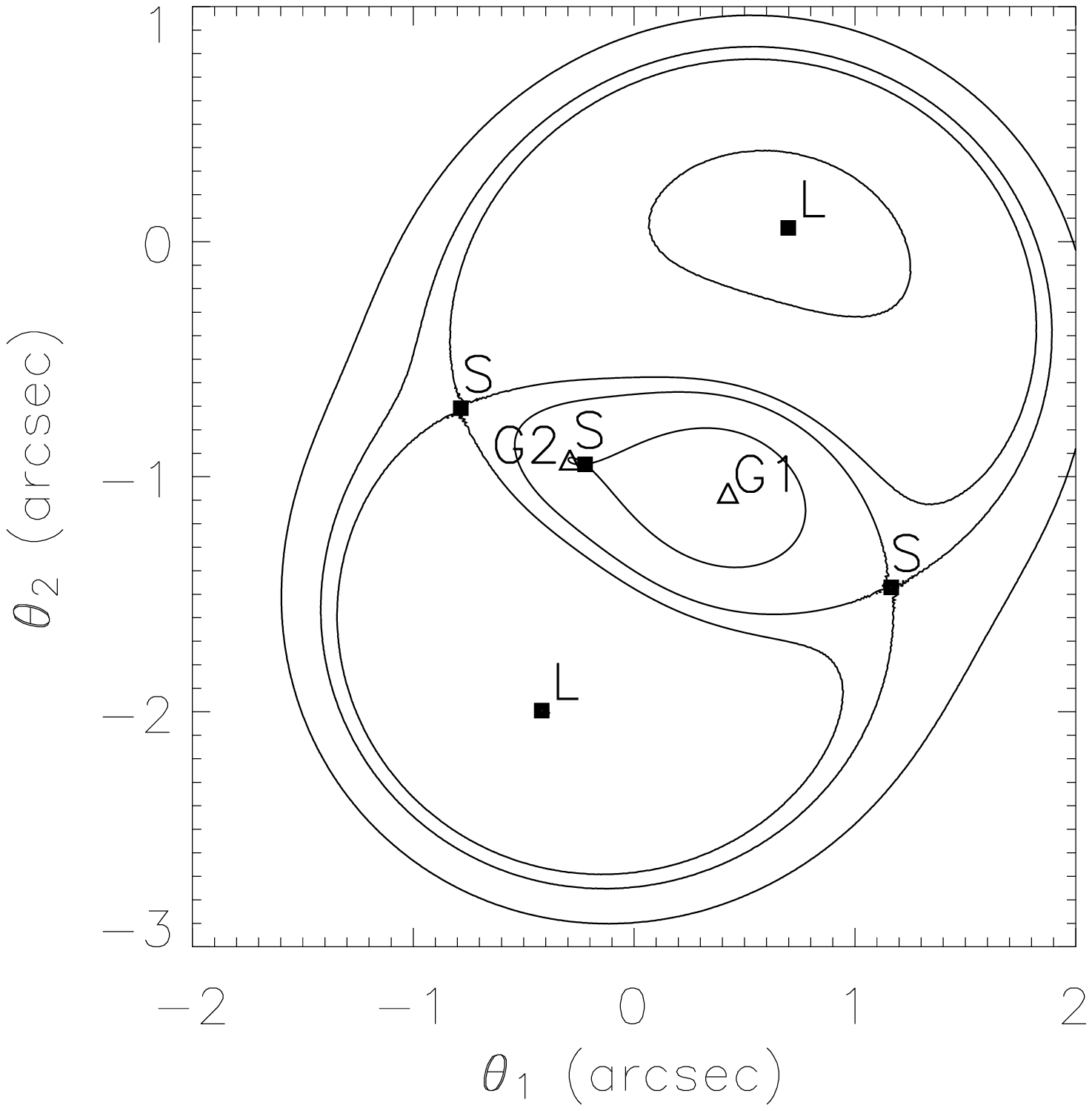}
\caption{\label{fig:imageTD2}  Left panel: source is near the centre within the astroid caustic of the B1608+656 SPLE1+D(isotropic) model in \citet{K03}.  Middle panel: the corresponding five images (A, B, C, D, and E), the lens galaxy positions (G1 and G2) indicated by open triangles, and the critical curves.  Right panel: crucial time delay contours for demonstrating Fermat's principle.  The time delay at each image position is a minimum (L for ``low'') or a saddle (S).  The scales on the source plane and image plane are different due to magnification of the images.}  
\end{figure*}

Fig.~\ref{fig:imageTD34} shows the image locations and the time delay contours as the source moves across a fold from within the caustic.  As the source approaches a fold, two of the images (B and C for the upper fold of interest) that are separated by the critical curve come together.  When the source is on the fold, the two images merge to become one at the corresponding point on the critical curve.  Finally, when the source moves across the fold, the merged image disappears.  The merging and disappearance of the two images can be explained using the lemniscate time delay contour (the saddle with two minima) in the right panels.  When the source approaches a fold, the time delay saddle of the lemniscate joins with one of its two associated local minima; after the source crosses the fold, only one time delay minimum remains.

\begin{figure*}
\includegraphics[width=56mm]{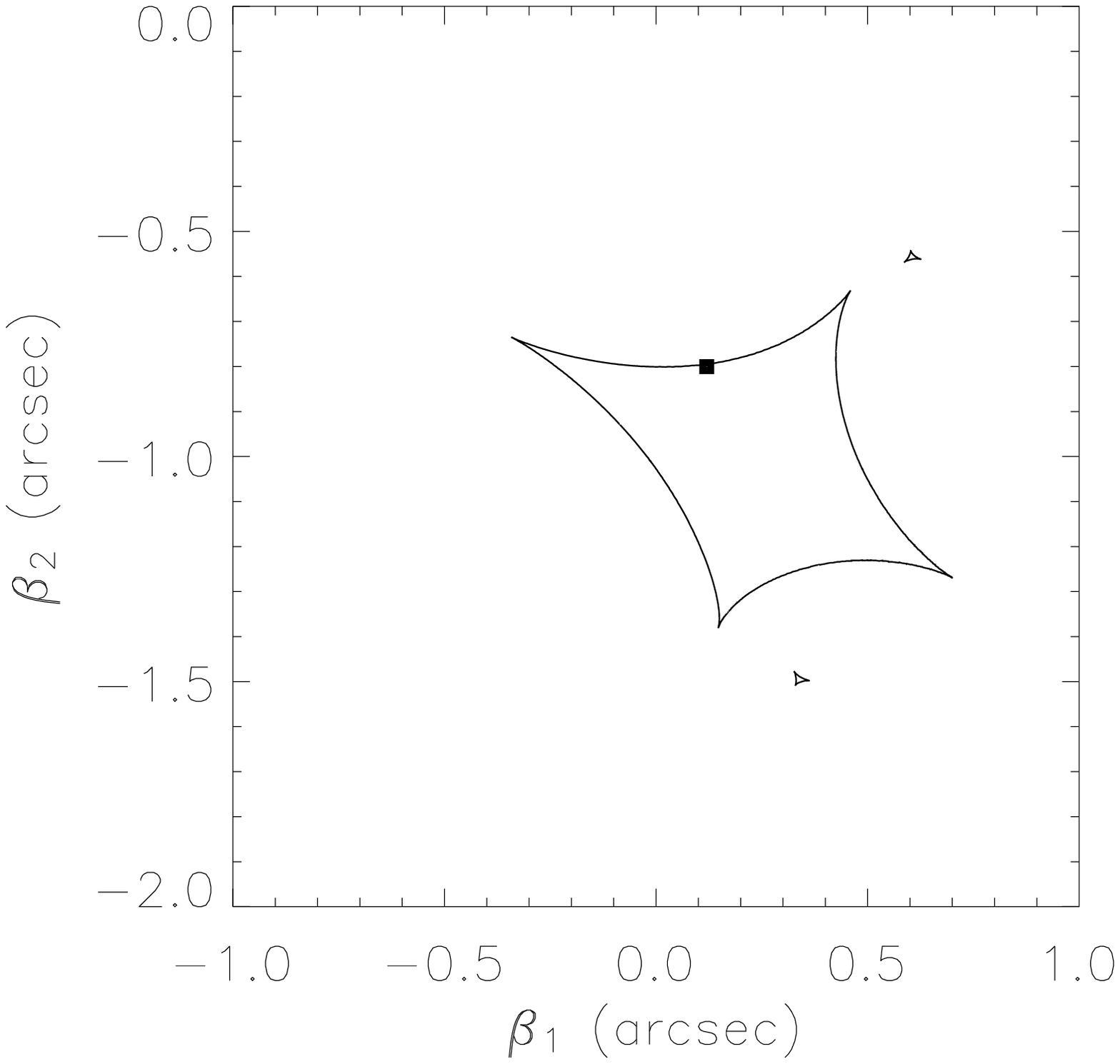}
\hspace{0.1in}
\includegraphics[width=56mm]{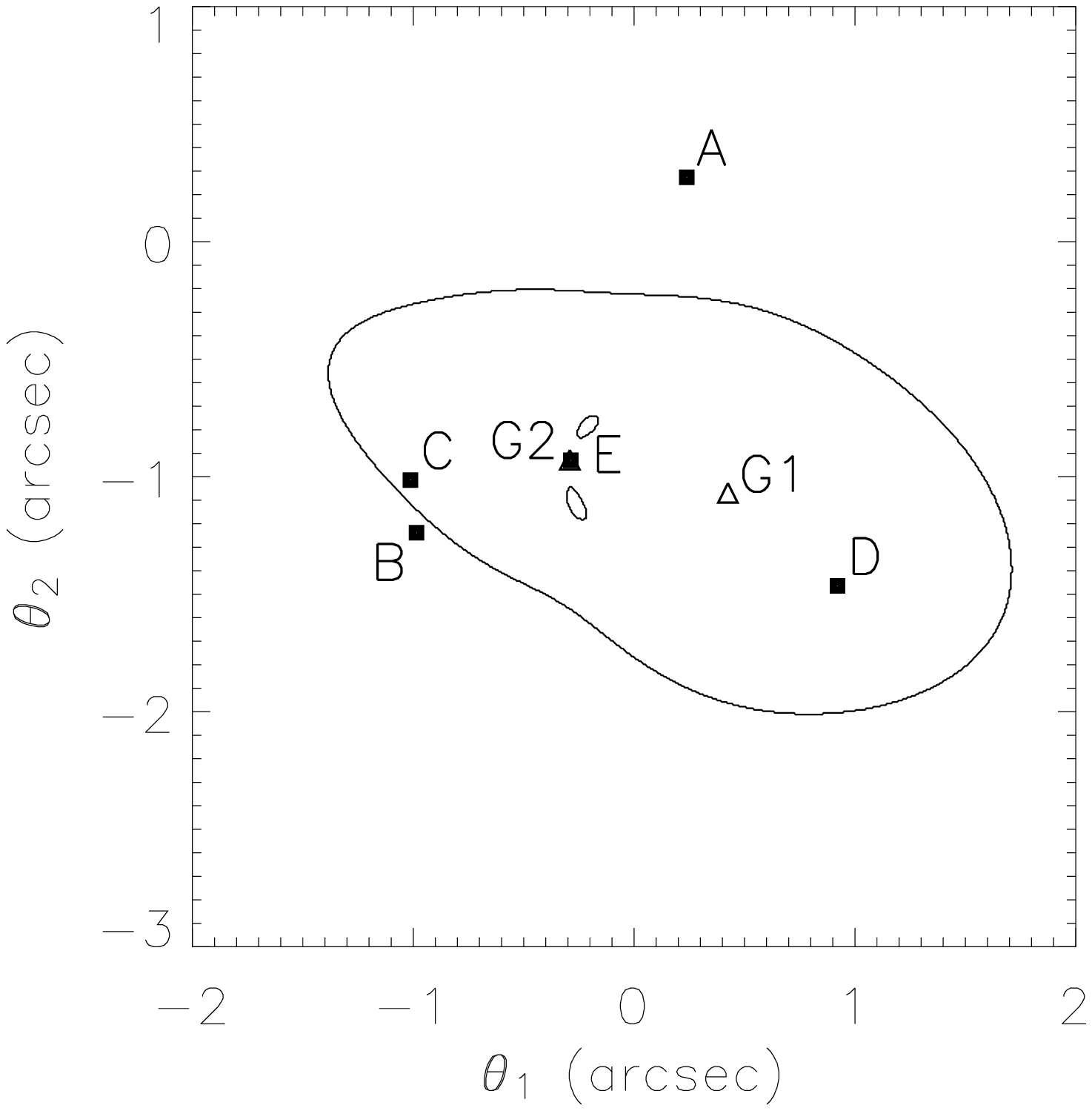}
\hspace{0.1in}
\includegraphics[width=56mm]{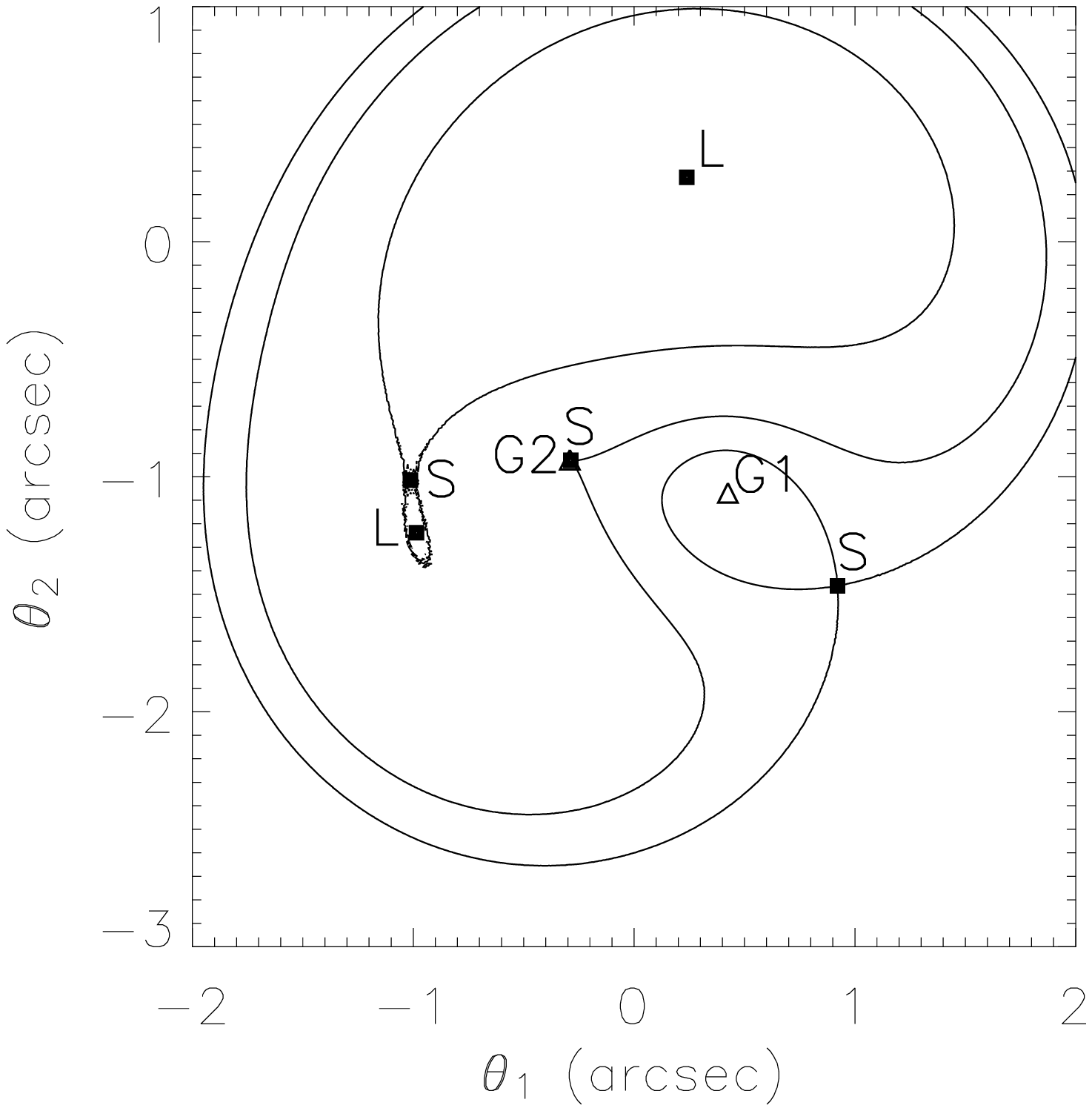}

\vspace{0.15in}
\includegraphics[width=56mm]{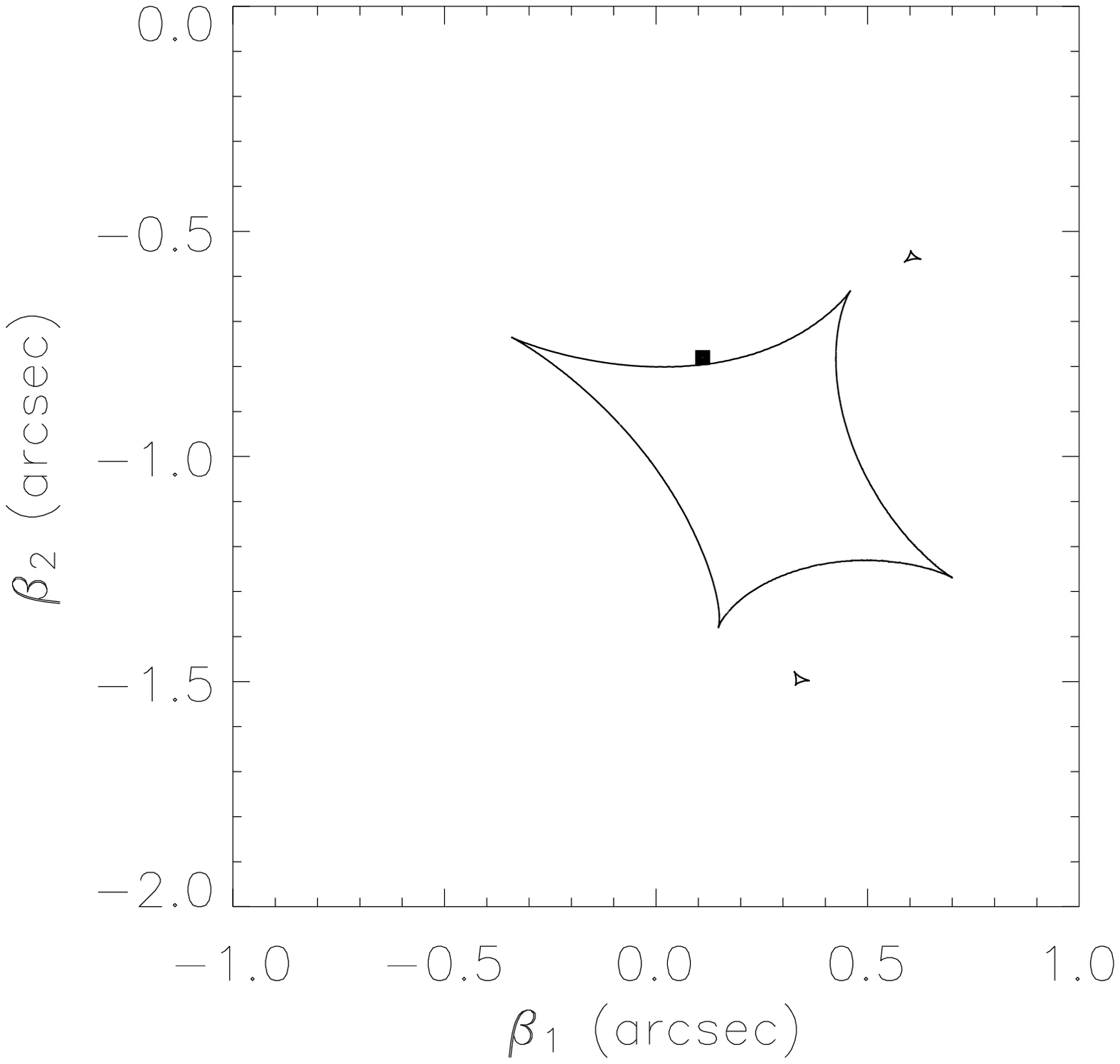}
\hspace{0.1in}
\includegraphics[width=56mm]{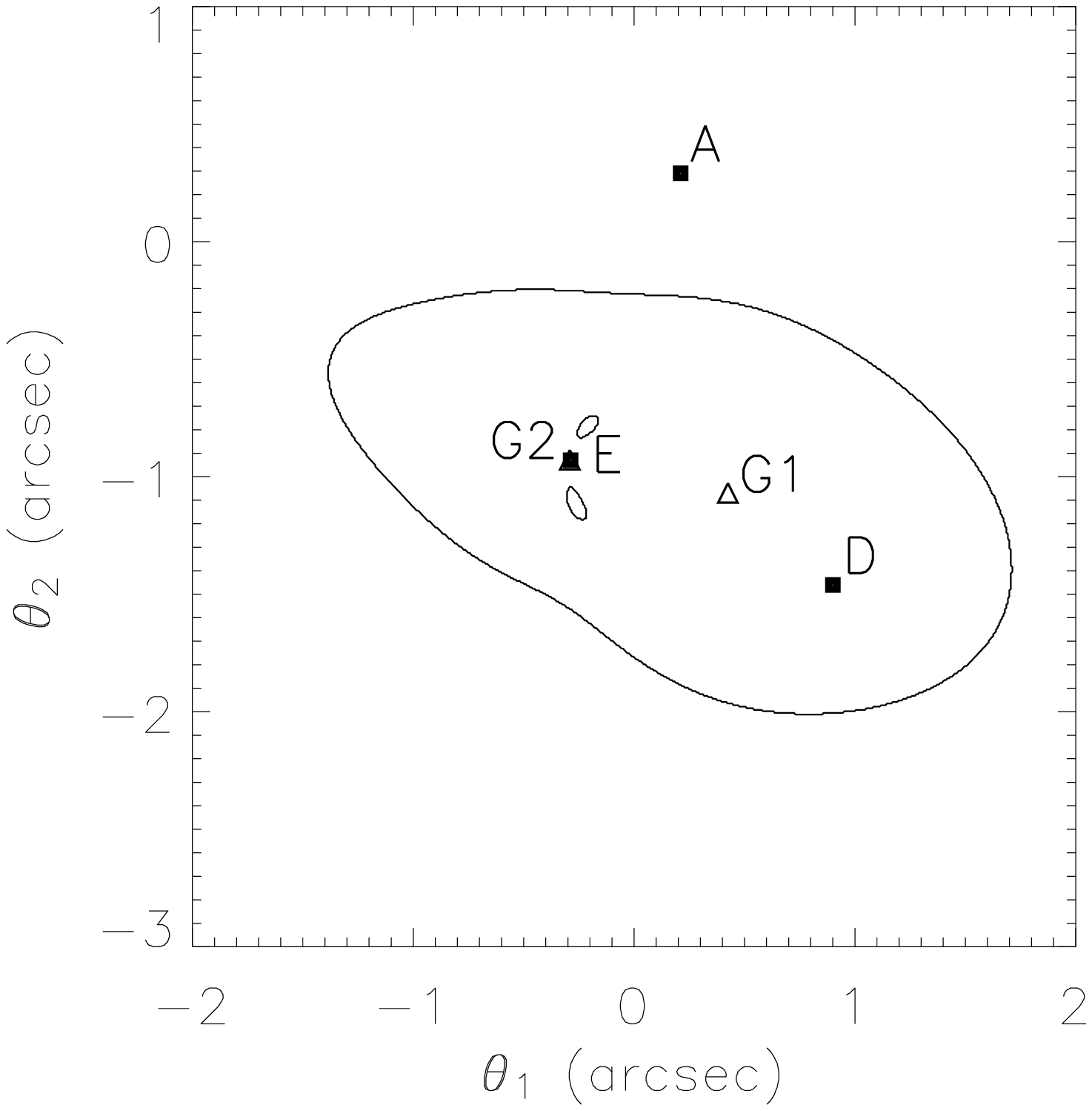}
\hspace{0.1in}
\includegraphics[width=56mm]{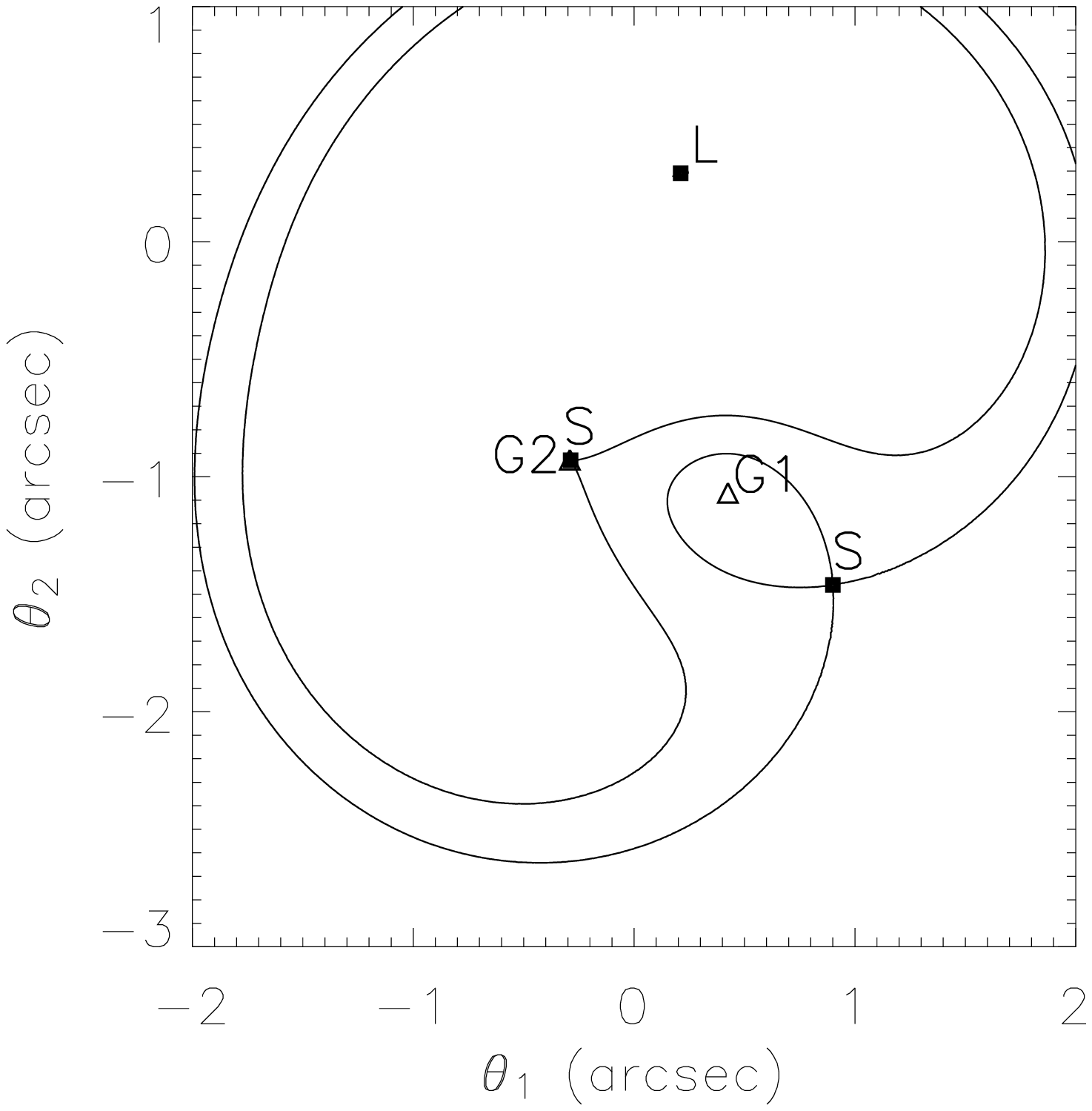}
\caption{\label{fig:imageTD34}  Left panels: source position displaced across a fold from inside (top) to outside (bottom) of the astroid caustic curve of B1608+656 SPLE1+D(isotropic) model.  Middle panels: image positions (A, B, C, D, and E) corresponding to the source positions shown in the left panels, lens galaxy positions (G1 and G2) indicated by open triangles, and the critical curves.  Right panels: corresponding time delay contours.  Letter L (for low) or S at each image location represents a time delay minimum or saddle, respectively.}  
\end{figure*}

Fig.~\ref{fig:imageTD78} shows the image locations and the arrival time delay contours as the source moves from within the astroid caustic across a cusp in a direction that is roughly along the semi-major axis of the lens distribution.  As the source approaches the cusp, three of the images (A, B, and C in this case) come together.  Two images (A and B) are outside and one image (C) is inside the critical curve.  When the source is on the cusp, the three images become one on the critical curve.  Finally, when the source moves across the cusp, one image remains outside the critical curve.  (We label the remaining image by the one that comes alphabetically first among the three merging images.)  The time delay contours in the right panels depict this behaviour: the time delay saddle of a lemniscate merges simultaneously with both of its two minima and leaves a single minimum in the end.

\begin{figure*}
\includegraphics[width=56mm]{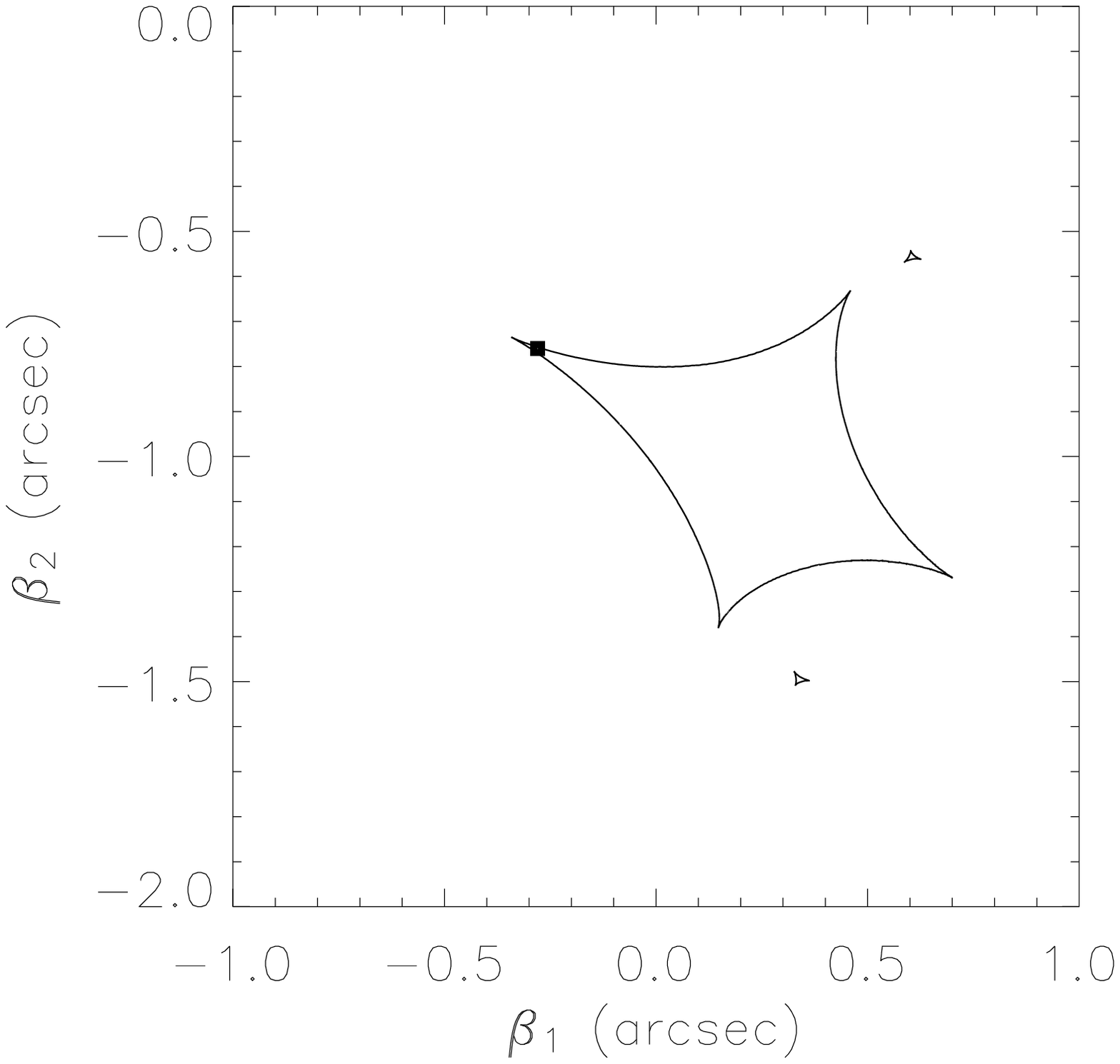}
\hspace{0.1in}
\includegraphics[width=56mm]{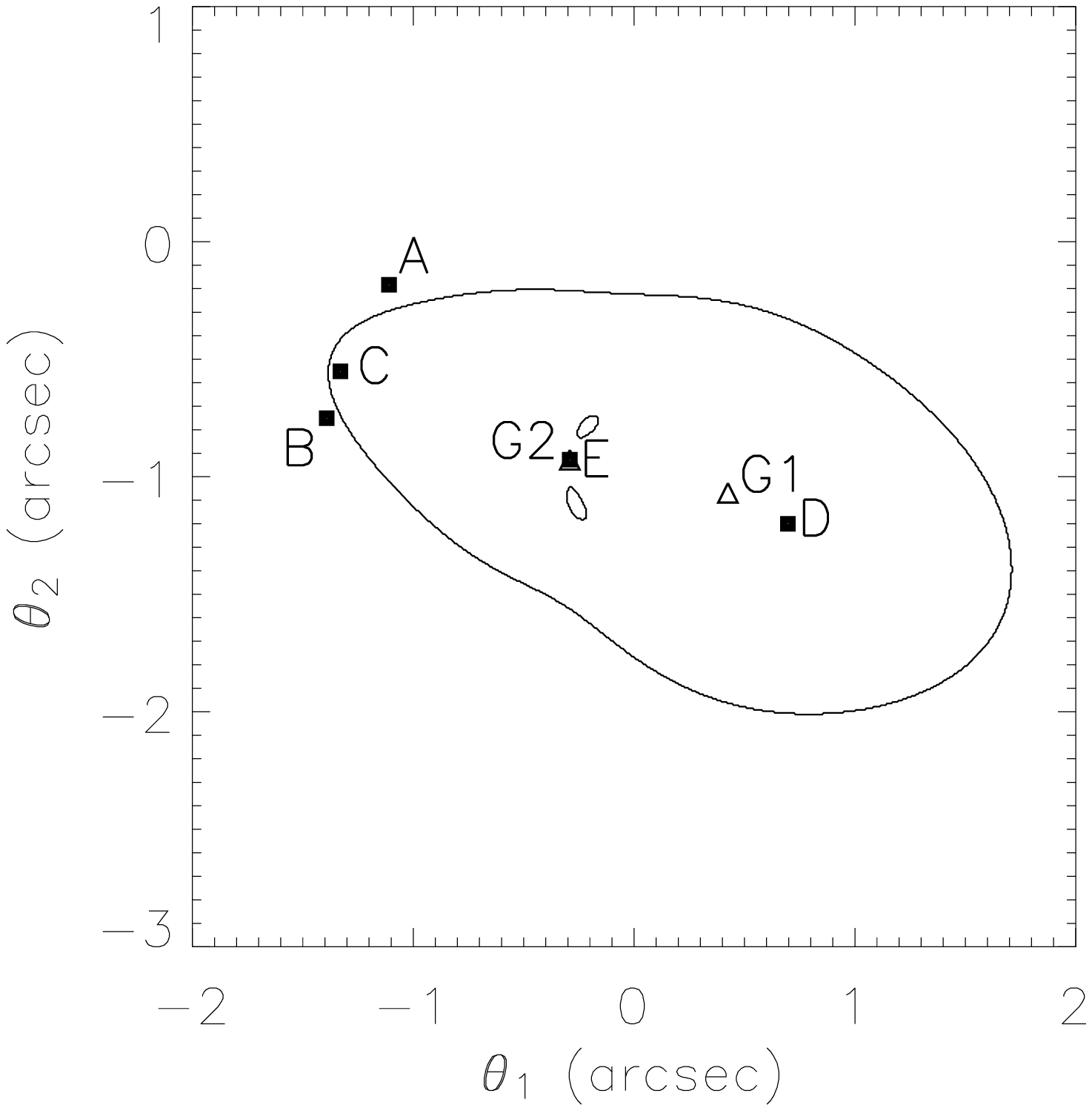}
\hspace{0.1in}
\includegraphics[width=56mm]{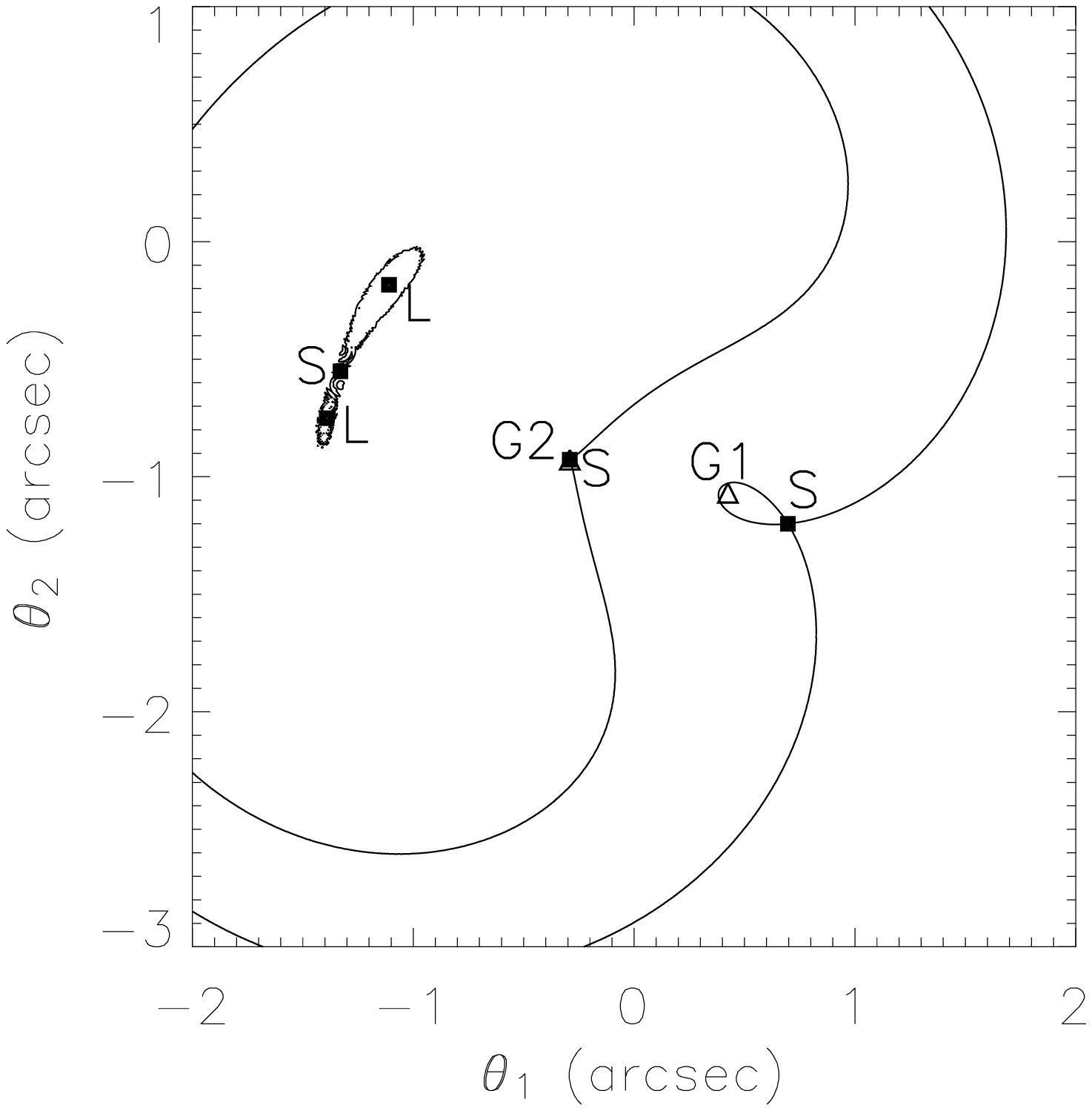}

\vspace{0.15in}
\includegraphics[width=56mm]{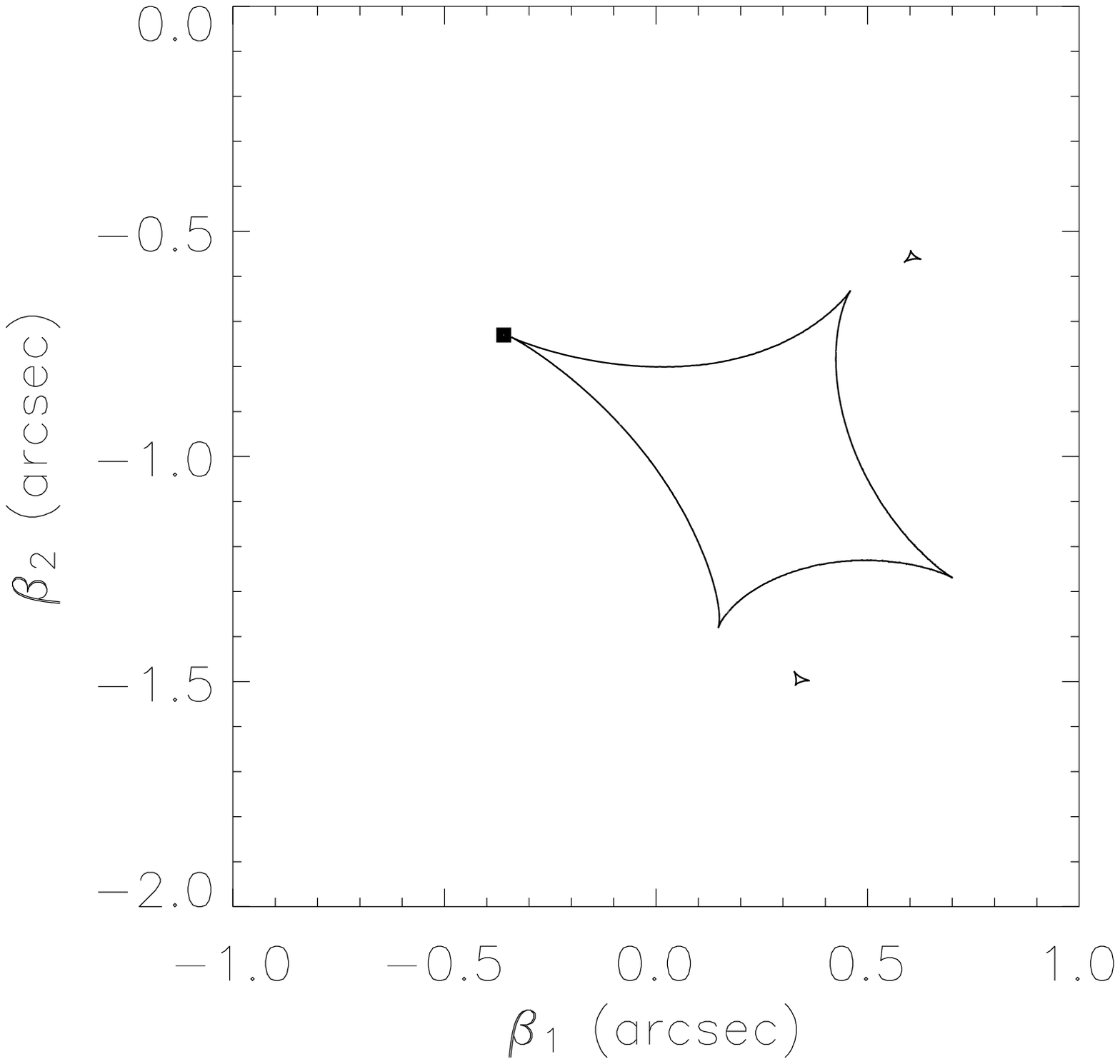}
\hspace{0.1in}
\includegraphics[width=56mm]{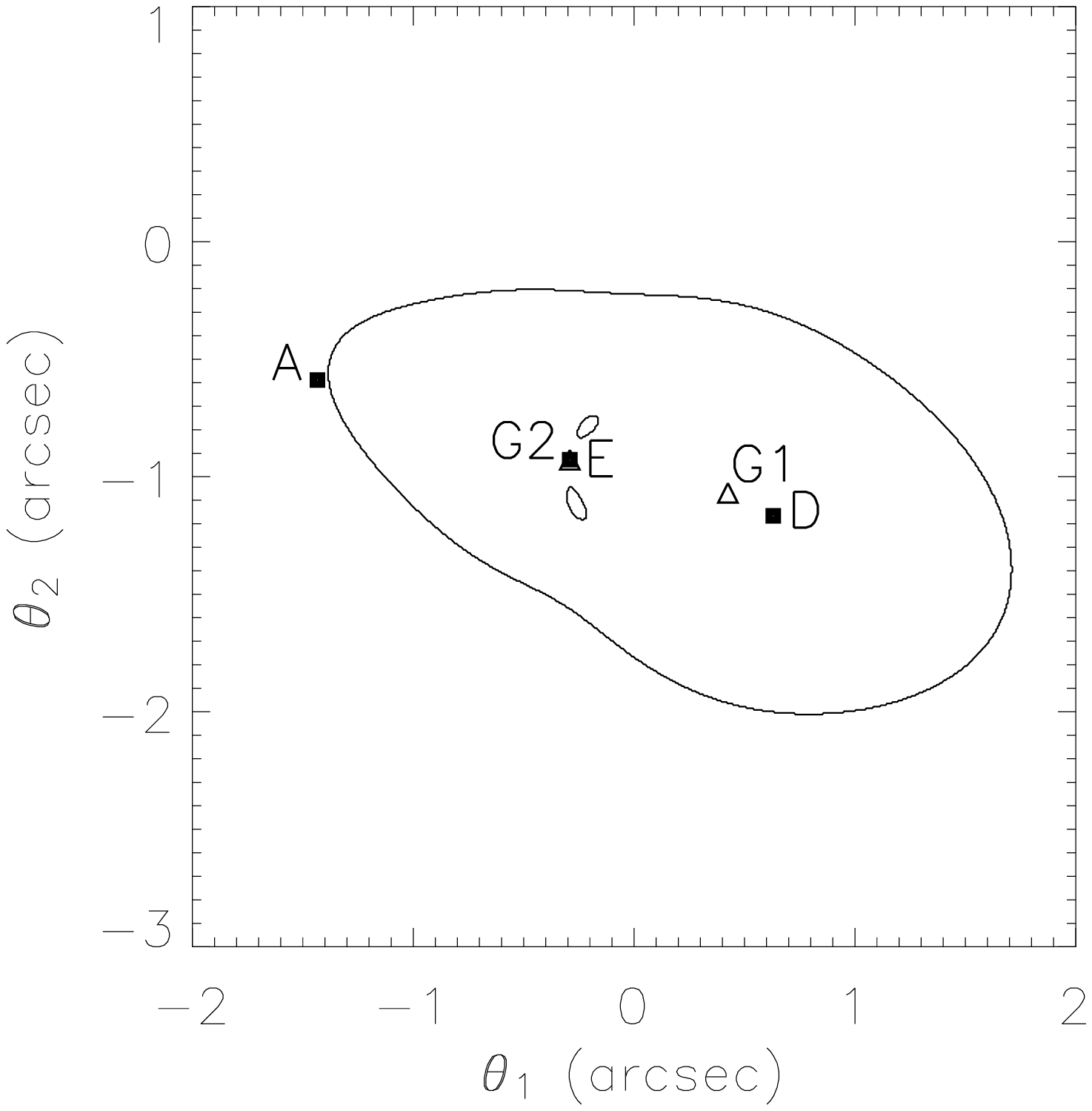}
\hspace{0.1in}
\includegraphics[width=56mm]{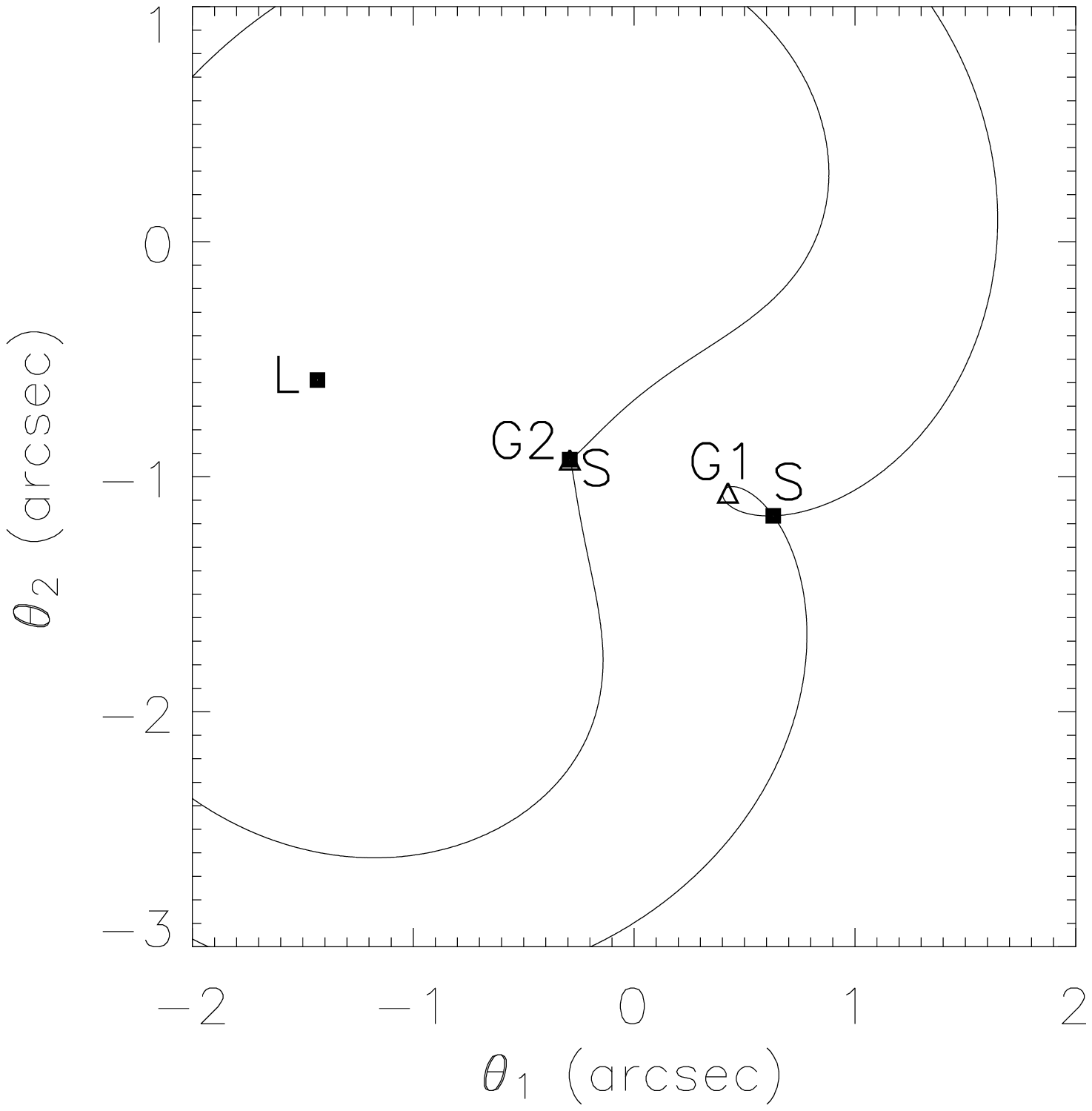}
\caption{\label{fig:imageTD78}  Left panels: source position displaced across a cusp approximately along the semi-major axis from inside (top) to outside (bottom) of the astroid caustic curve of B1608+656 SPLE1+D(isotropic) model.  Middle panels: image positions (A, B, C, D, and E) corresponding to the source positions shown in the left panels, lens galaxy positions (G1 and G2) indicated by open triangles, and the critical curves.  Right panels: corresponding time delay contours.  Letter L (for low) or S at each image location represents a time delay minimum or saddle, respectively.}
\end{figure*}

Fig.~\ref{fig:imageTD56} is similar to Fig.~\ref{fig:imageTD78} but with the source displacing toward a cusp that is roughly along the semi-minor axis of lens distribution. The three merging images now have one image (B) outside and two images (C and D) inside the critical curve (shown in middle panels).  In terms of the time delay contours (right panels), this corresponds to the simultaneous merging of the saddle of the lemniscate with one of its minima and with the saddle of the enclosing lima\c con, leaving only the lima\c con saddle in the end.

\begin{figure*}
\includegraphics[width=56mm]{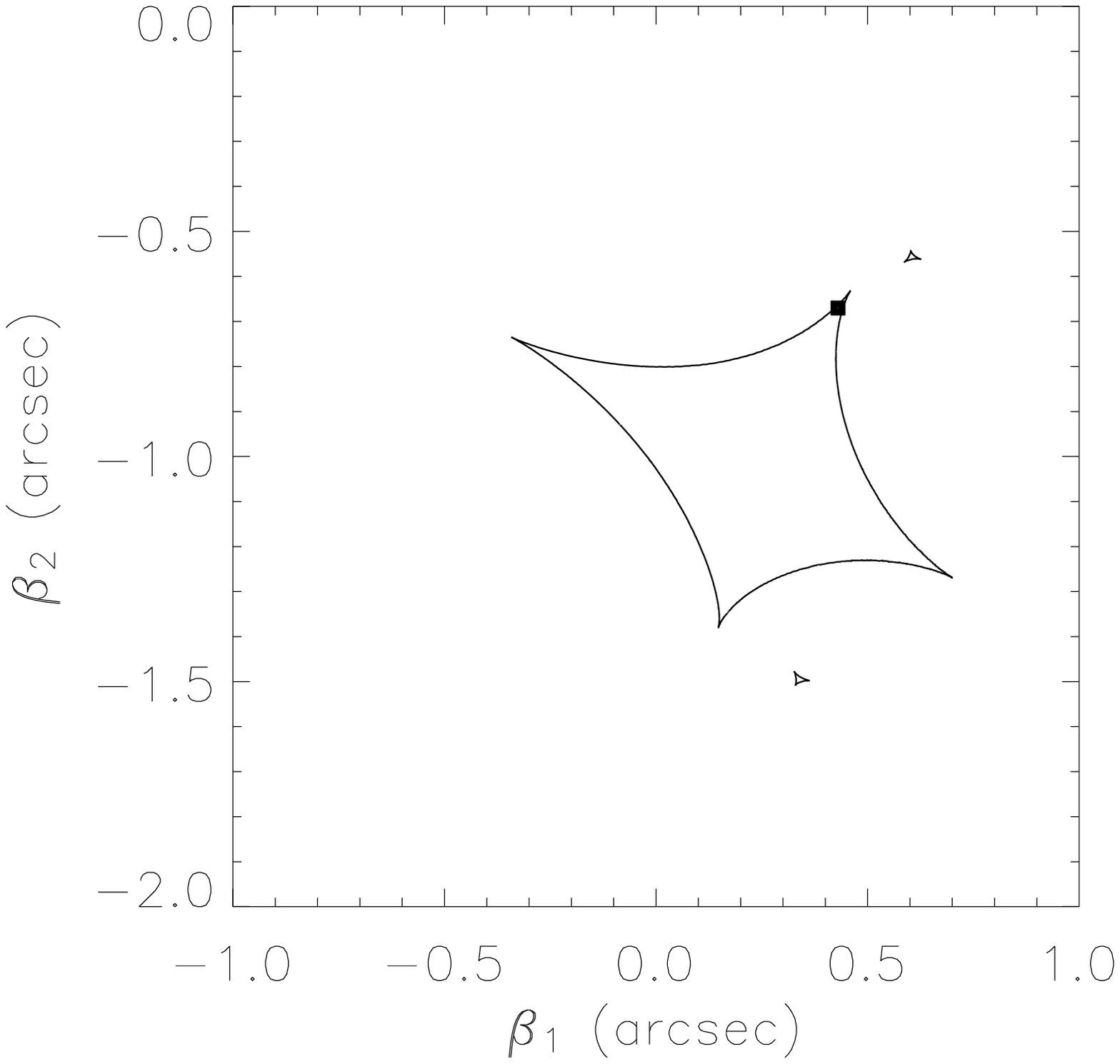}
\hspace{0.1in}
\includegraphics[width=56mm]{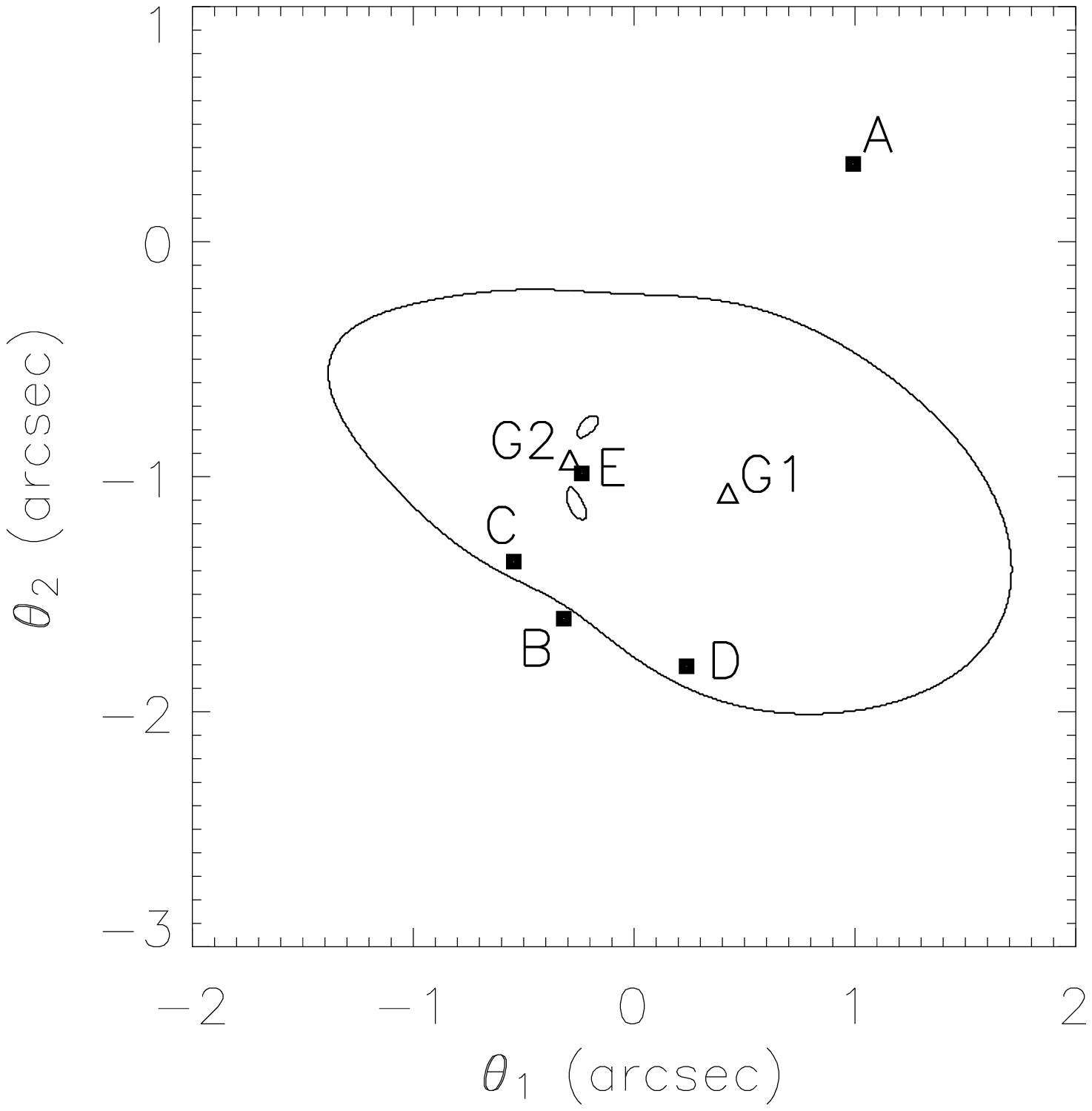}
\hspace{0.1in}
\includegraphics[width=56mm]{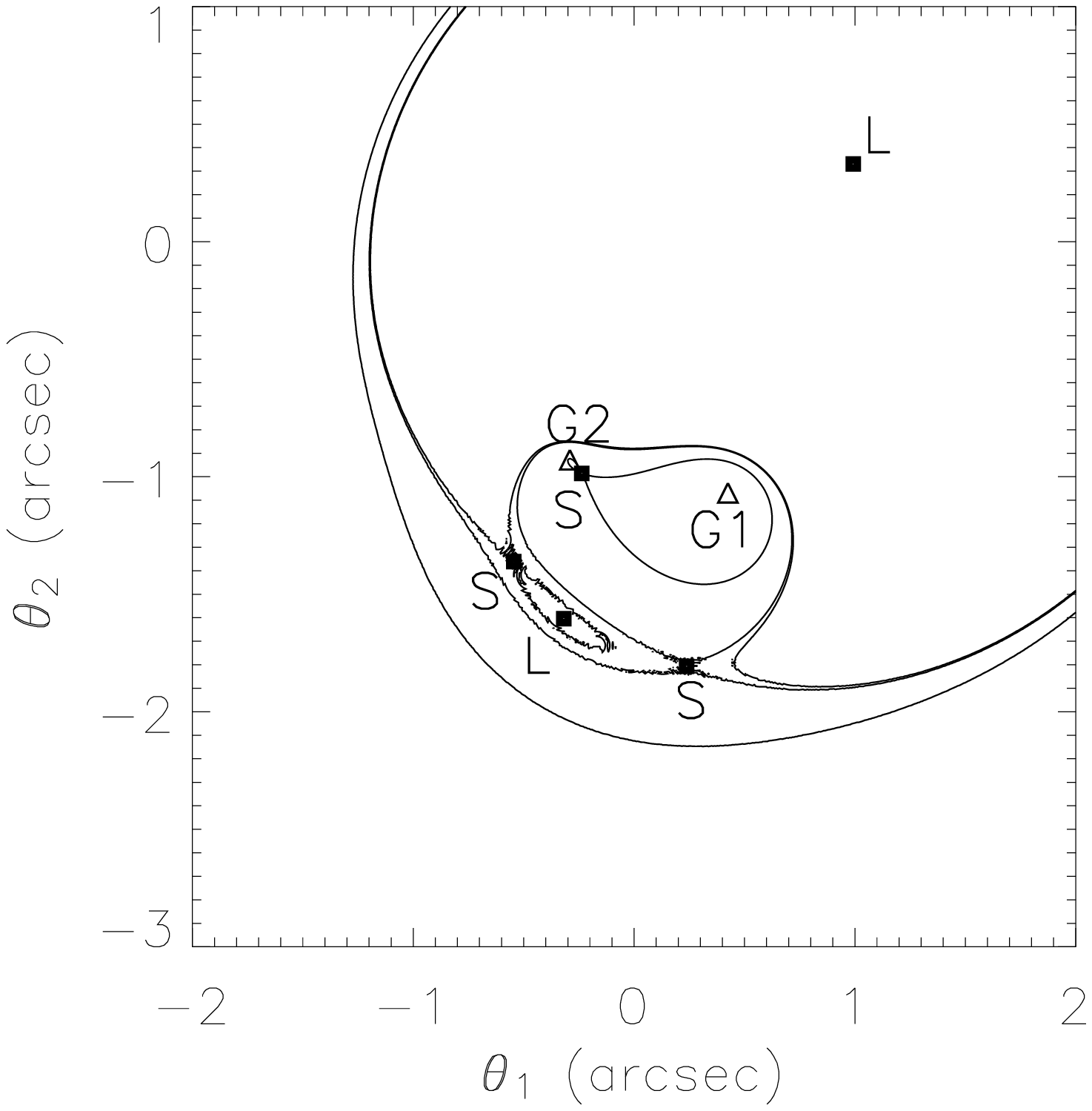}

\vspace{0.15in}
\includegraphics[width=56mm]{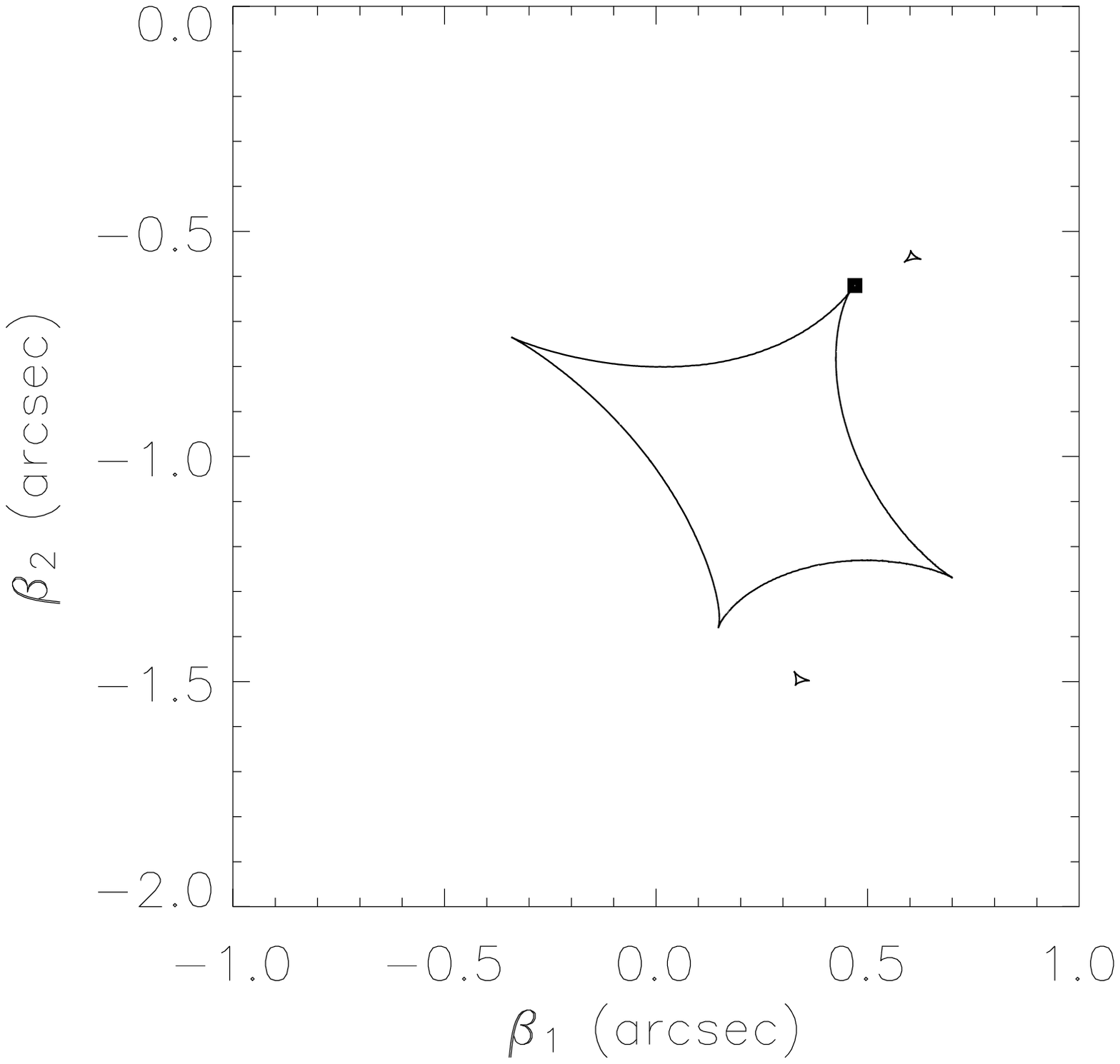}
\hspace{0.1in}
\includegraphics[width=56mm]{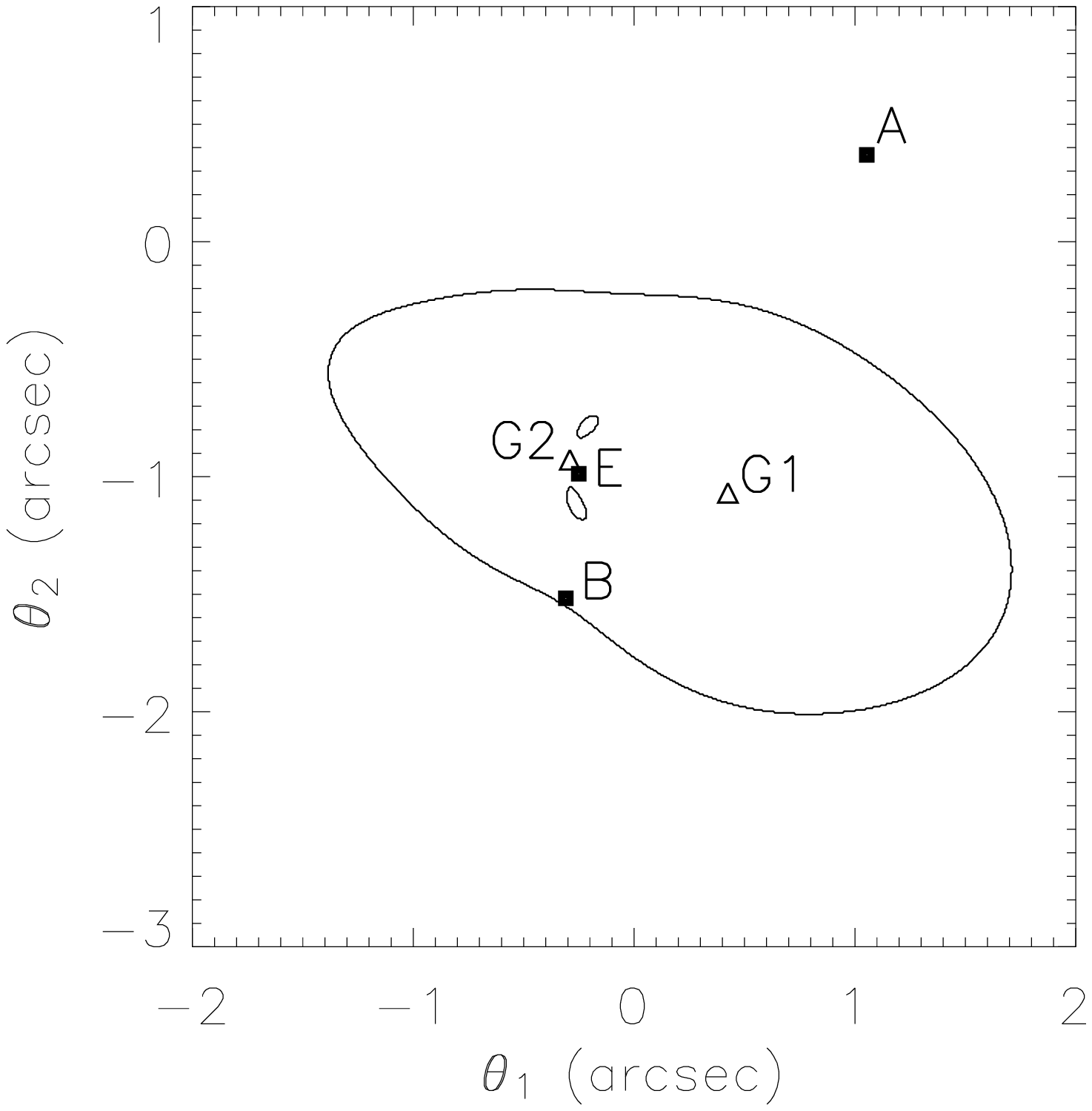}
\hspace{0.1in}
\includegraphics[width=56mm]{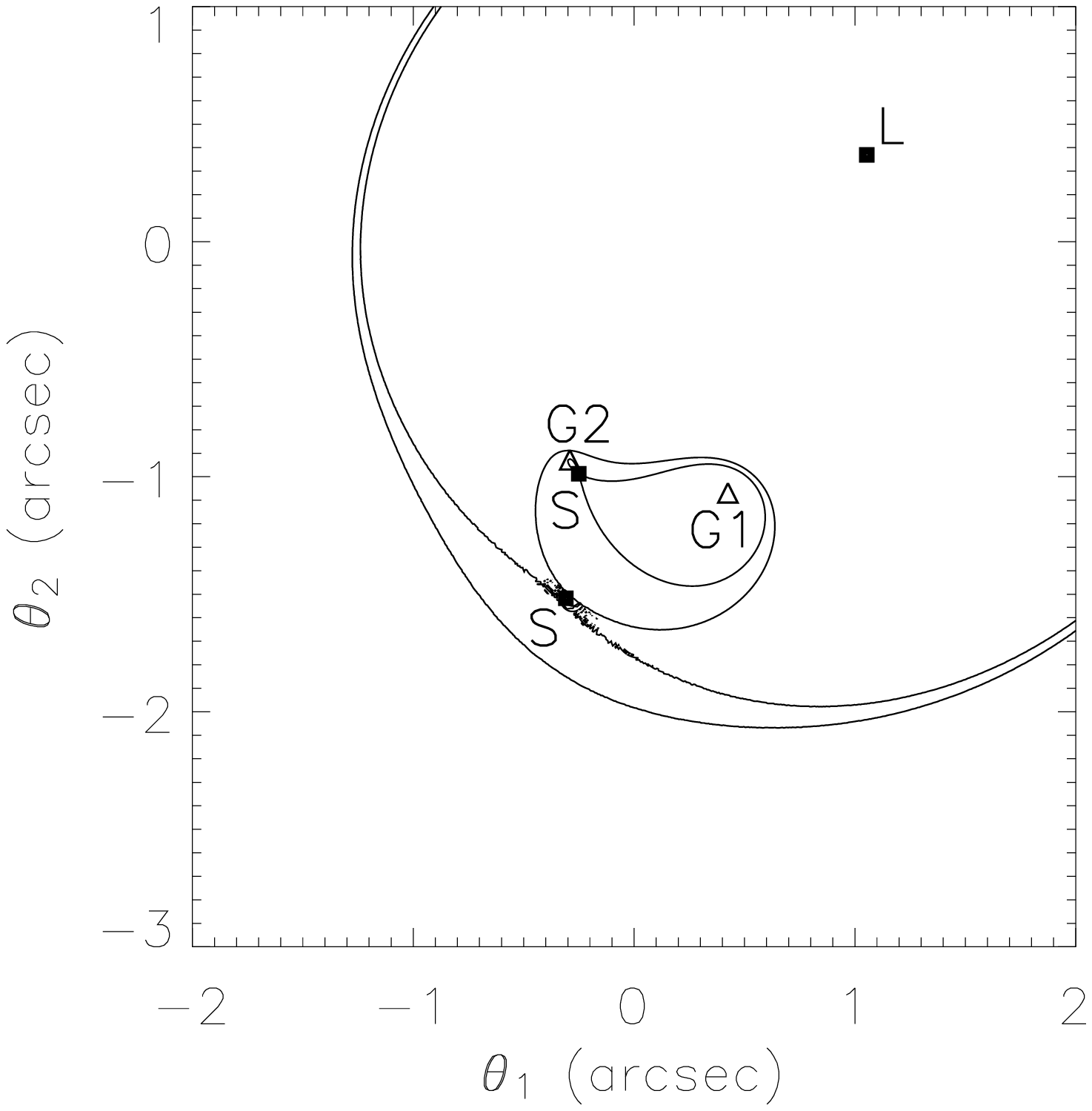}
\caption{\label{fig:imageTD56}  Left panels: source position displaced across a cusp approximately along the semi-minor axis from inside (top) to outside (bottom) of the astroid caustic curve of B1608+656 SPLE1+D(isotropic) model.  Middle panels: image positions (A, B, C, D, and E) corresponding to the source positions shown in the left panels, lens galaxy positions (G1 and G2) indicated by open triangles, and the critical curves.  Right panels: corresponding time delay contours.  Letter L (for low) or S at each image location represents a time delay minimum or saddle, respectively.}
\end{figure*}

\subsubsection{Inner and outer limits}

The movements of the image locations shown in Figs.~\ref{fig:imageTD2} to \ref{fig:imageTD56} allow us to define \textit{limit curves} \citep{BN86}.  Consider moving a hypothetical point source on the caustic curve.  As the source traces around the folds of the caustic, the two non-merging images trace out the limit curves.  For the astroid, the non-merging image inside the critical curve is on the \textit{inner limit} and the image outside the critical curve is on the \textit{outer limit}.  For the deltoids, both non-merging images are outside the corresponding critical curves.  The deltoids thus have only outer limits composed of two images and no inner limits.  Fig.~\ref{fig:limitCurv} is the plot of the limit curves for the SPLE1+D(isotropic) model in \citet{K03}.  The inner limit and outer limit for the astroid are shown in green and orange, respectively.  The outer limit for the deltoids are shown in cyan.  

We focus only on the limit curves of the astroid since they are typical for elliptical lens mass distributions.  Both the inner and the outer limits are tangent to the critical curve twice, corresponding to source placement at the cusps of the caustic.  The limit curves mark the boundary of the region containing four images.

\begin{figure}
\includegraphics[width=84mm]{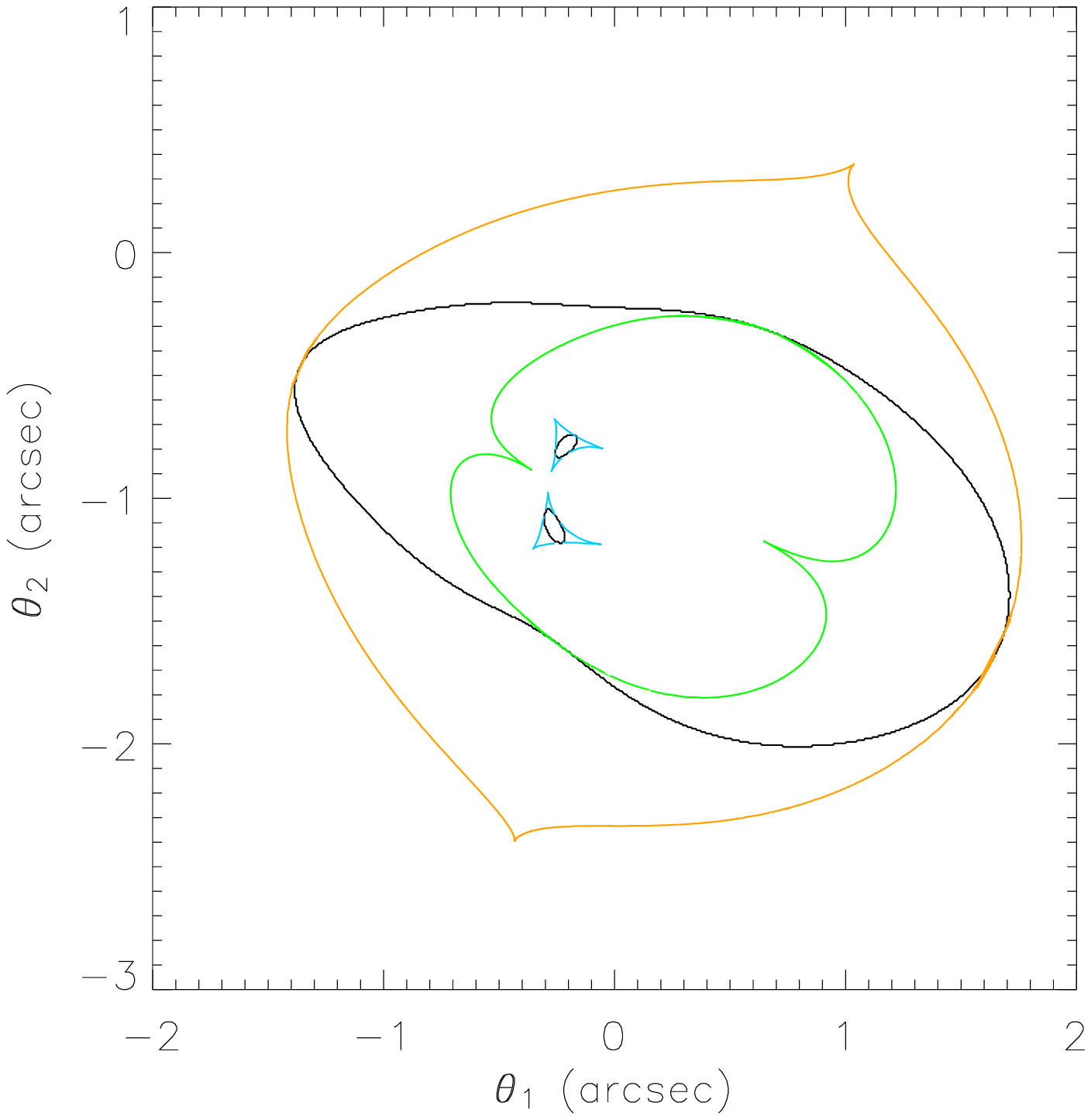}
\caption{\label{fig:limitCurv} The limit curves are plotted with the critical curves (in black) for the SPLE1+D(isotropic) model of B1608+656.  The orange (green) curve is the outer (inner) limit associated with the astroid.  The limit curves are each tangent to the critical curve of the astroid twice.  The cyan curves are the outer limits of the deltoids.}  
\end{figure}

\subsubsection{Isophotal separatrices}

An isophote is an intensity contour.  We assume the source intensity distribution has a single maximum with nested, non-crossing contours.  An \textit{isophotal separatrix} on the image plane corresponds to a source intensity contour that is tangent to the caustic curve.  The isophotes must cross at the critical curve and be tangent to the limit curves as we explain below.

Consider an extended elliptical source intensity distribution centred at $(\beta_{s1},\beta_{s2})=(0.088,-1.069)$ with an axis ratio of 0.634 and a semi-major axis position angle of 22.1 degrees\footnote{This source model differs from the \citet{K03} source model in the position angle, but the difference is irrelevant for the purpose of describing isophotal separatrices.}.  The left panel in Fig.~\ref{fig:crossingIso} shows four coloured intensity contours of the extended source.  The two intermediate isophotes are very close together (light blue and dark blue).  The right panel in Fig.~\ref{fig:crossingIso} shows the mapped isophotes (same colours) with the critical curves (black) and limit curves (red).  Each coloured set of isophotes must intersect at the critical curve and be tangent to the inner and outer limit.  This is shown most clearly by the purple isophotes that consist of a lemniscate (separatrix) with two elliptical satellite isophotes on the image plane.  The lemniscate isophote must cross at the critical curve, and the two satellite isophotes must each be tangent to either the inner or the outer limit.  

To explain the crossing and tangency conditions, let us consider the purple isophotes in detail.  The crossing point of the lemniscate on the critical curve corresponds to the tangency point of the source isophote to the astroid caustic curve.  Recall from section 2.2.2 that two of the four images of a hypothetical point source merge on the critical curve as the source moves across the fold from within.  Therefore, a segment of the source isophote to either side of the caustic tangency point will map to two segments on the image plane, one inside and one outside the critical curve, that connect at the critical curve.  The entire source isophote that is within the caustic will thus correspond to a lemniscate crossing the critical curve on the image plane with one lobe inside and one lobe outside the critical curve.  The tangencies of the image isophotes to the limit curves can be understood based on the definition of limit curves, which are the inner and outer boundaries of the four-image region that are marked by the two non-merging images as a hypothetical source traces around the caustic.  The two satellite isophotes correspond to image isophotes traced by the two non-merging images that must touch the inner and outer limits when the source isophote is tangent to the caustic.  Since the inner and outer limits are the four-image boundaries, the touchings of the satellite isophotes to the limit curves become tangencies.  Similar reasoning applies to the crossings and tangencies of the other three sets of isophotes.

The crossing of the isophotes at the critical curves and the tangency of the isophotes to the limit curves provide qualitative tests on how good a lens model is.

\begin{figure*}
\includegraphics[width=84mm]{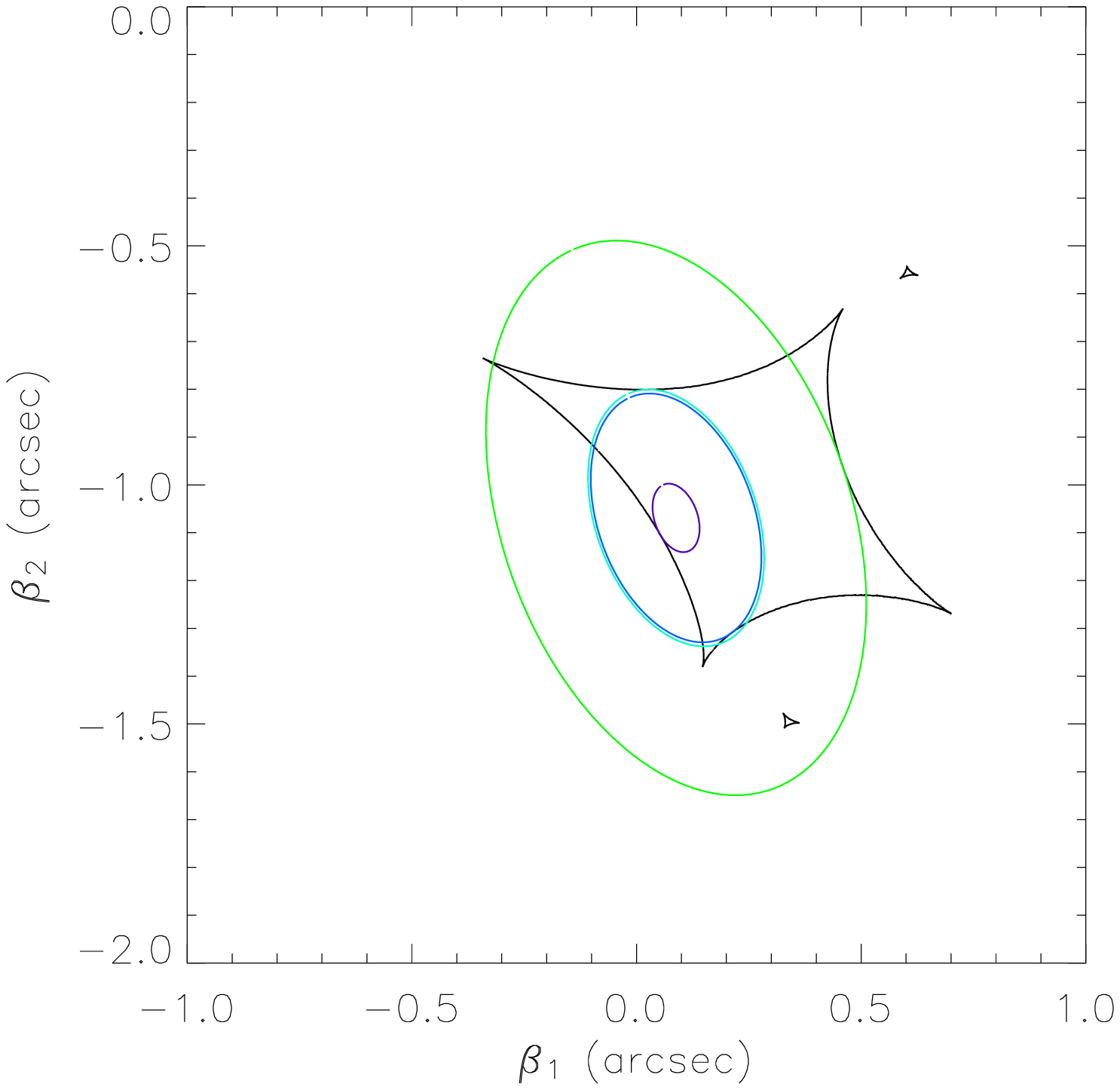}
\hspace{0.2in}
\includegraphics[width=84mm]{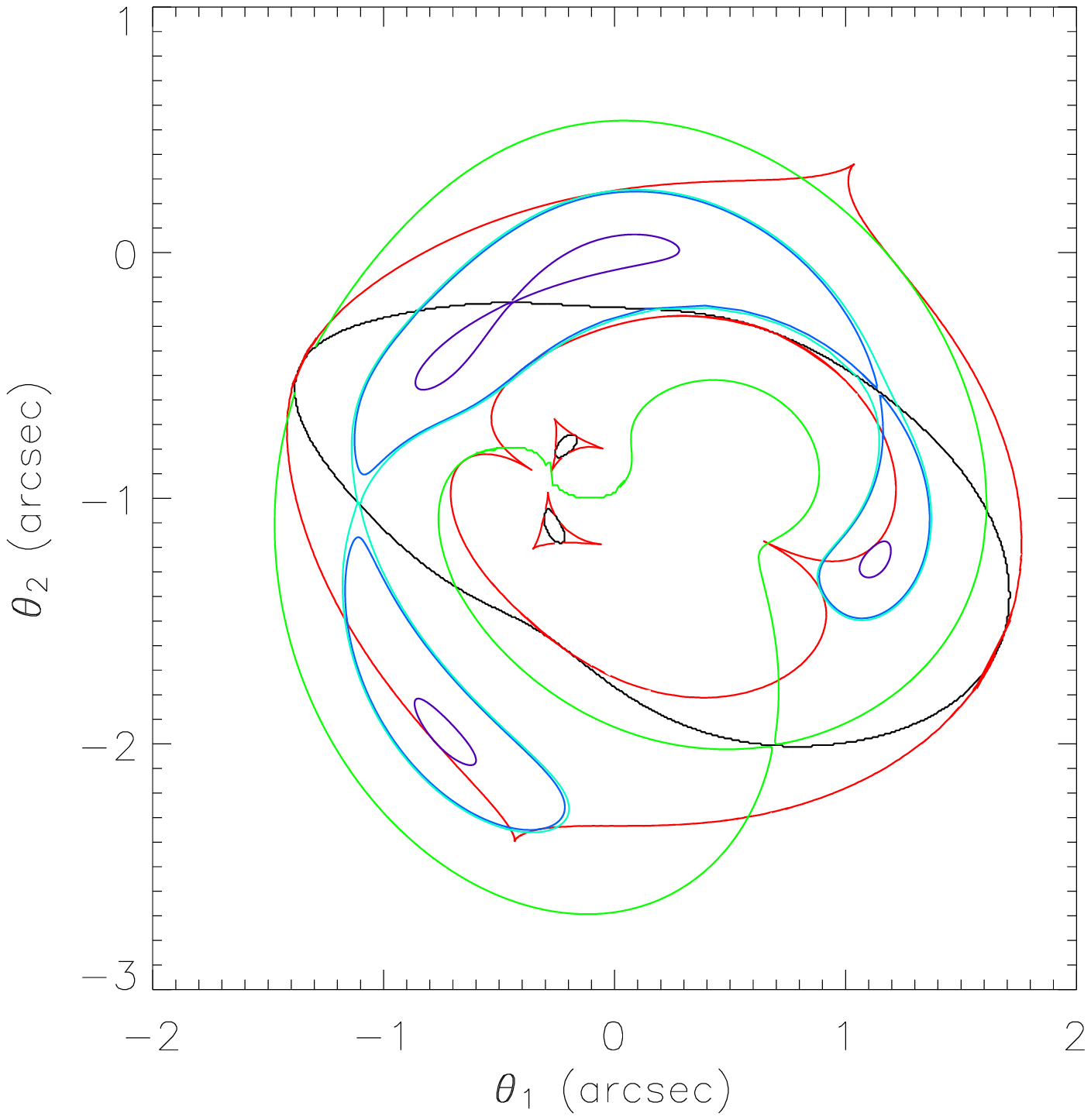}
\caption{\label{fig:crossingIso} Left panel: four isophotes (in colours) of an extended source intensity distribution that are tangent to the astroid caustic curve (black).  Right panel: the mapping of the isophotes in the left panel with the critical curve (black) and limit curves (red).  These isophotes must cross at the critical curve and their satellite isophotes must be tangent to the limit curves.}
\end{figure*}

\subsubsection{Observational Data}

We use the result of section 2.2.4 to qualitatively test the SPLE1+D(isotropic) model in \citet{K03} by superimposing the critical and limit curves of the model on the intensity contours of the observational data.  Fig.~\ref{fig:F814Curves} shows the isophotal separatrices (in black in various line styles) of the deconvolved residual \textit{HST}/F814W image of B1608+656 \citep{K03} with the critical curves (red) and limit curves (green, orange, cyan).  We check the crossing and tangency conditions for each of the four sets of isophotal separatrices, using Fig.~\ref{fig:crossingIso} as a guide for the approximate crossing and tangency locations.  For the dashed isophotes, the conditions for the crossing of the separatrix at the critical curve and the tangency to the limit curves are violated.  For the solid isophotes, the crossing at $(\theta_1, \theta_2) \sim(-0.8,-1.1)$ is not at the critical curve, but the tangency requirements at $\sim(0.9,-1.4)$ and $ \sim(0.4,0.3)$ are satisfied within the noise.  For the dotted isophotes, the crossing at $\sim (0.7,-1.9)$ is at the critical curve within the noise, but the isophotes near $\sim (-0.5,-0.9)$ and $\sim (1.1,0.2)$ are not tangent to the limit curves.  Lastly, for the long-dashed isophotes, the crossing at $\sim(1.3,-0.6)$ is on the critical curve, and the isophotes near $\sim(-0.5,-0.6)$ and $\sim(-0.3, -2.4)$ are tangent to the limit curves, within the noise.  Therefore, the SPLE1+D(isotropic) model proposed by \citet{K03} satisfies the crossing and tangency conditions stated in section 2.2.4 for some, but not all, of the isophotal separatrices.  As a result, the model proposed by \citet{K03} must not represent the true lens potential of the system, especially in the regions where the crossings and tangencies fail.  Recall that we need an accurate model of the lens potential to calculate the Hubble constant.  In the next section, we examine a method of potential correction.

\begin{figure}
\includegraphics[width=84mm]{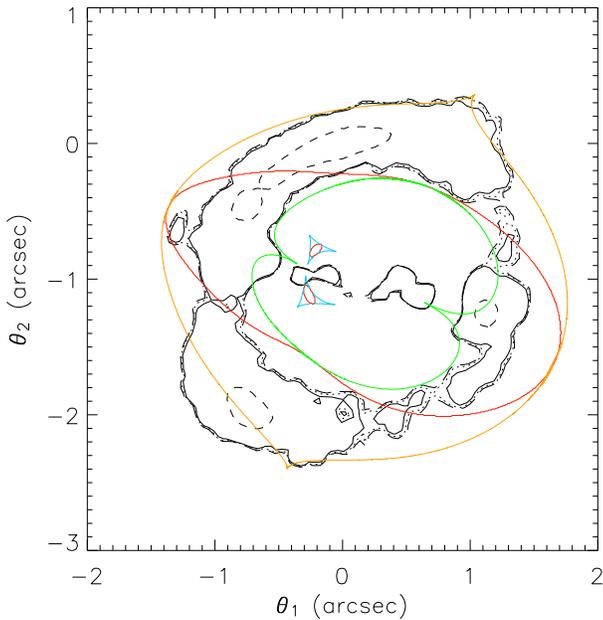}
\caption{\label{fig:F814Curves} The deconvolved residual \textit{HST}/F814W image of B1608+656 \citep{K03}.  The isophotal separatrices (in black in various line styles) are shown with the critical curves (red) and limit curves (green, orange, and cyan) of the SPLE1+D(isotropic) model in \citet{K03}.  Some of the isophotal separatrices are not intersecting at the critical curve of the model and some of the satellite isophotes are not tangent to the limit curves of the model.}
\end{figure}

\section{Potential Reconstruction}

\subsection{Theory of potential reconstruction}

The method of potential reconstruction was first suggested by \citet{BSK01}.  Following the notation in section 2.1, let $I(\bmath{\theta})$ be the observed image intensity of a gravitational lens system with an extended source.  For a given potential model, $\psi(\bmath{\theta})$, one can obtain the best-fitting source intensity distribution \citep{WD03}.  Let $I(\bmath{\beta})$ be the source intensity translated to the image plane via the potential model, $\psi({\bmath{\theta}})$.  We define the intensity deficit on the image plane by
\be \label{eq:dI} 
\delta I(\bmath{\theta}) = I(\bmath{\theta}) - I(\bmath{\beta}).
\ee
The intensity deficit is zero everywhere with the true lens potential distribution.  

Consider a lens potential model that is perturbed from the true potential, $\psi_{0}(\bmath{\theta})$, by $\delta \psi(\bmath{\theta})$:
\be \label{eq:potmodel}
\psi(\bmath{\theta}) = \psi_{0}(\bmath{\theta}) + \delta \psi(\bmath{\theta}).
\ee
We can correct the potential model perturbatively by solving for the perturbation $\delta \psi(\bmath{\theta})$.  For a given image (fixed $\bmath{\theta}$ and $I(\bmath{\theta})$), we can relate a change in position on the source plane, $\delta \bmath{\beta}$, to the potential perturbation using the lens equation (\ref{eq:lensEq}):
\be \label{eq:dSrPos}
\delta \bmath{\beta} = - \frac {\partial \delta \psi(\bmath{\theta})}{\partial \bmath{\theta}}.
\ee
Expanding $I(\bmath{\beta})$ to first order in $\delta \bmath{\beta}$ and using equation (\ref{eq:dSrPos}) in equation (\ref{eq:dI}), we obtain
\be \label{eq:pertEq}
\delta I(\bmath{\theta}) = -\frac{\partial I(\bmath{\beta})}{\partial \bmath{\beta}} \bmath{\cdot} \delta \bmath{\beta} = \frac{\partial I(\bmath{\beta})}{\partial \bmath{\beta}} \bmath{\cdot} \frac {\partial \delta \psi(\bmath{\theta})}{\partial \bmath{\theta}}.
\ee
The source intensity gradient $\frac{\partial I(\bmath{\beta})}{\partial \bmath{\beta}}$ implicitly depends on the potential model $\psi(\bmath{\theta})$ since the source position $\bmath{\beta}$ (where the gradient is evaluated) is related to $\psi(\bmath{\theta})$ via the lens equation (\ref{eq:lensEq}).  To first order, using the perturbed model $\psi(\bmath{\theta})$ is equivalent to using the true model $\psi_0(\bmath{\theta})$ in the evaluation of the source intensity gradient $\frac{\partial I(\bmath{\beta})}{\partial \bmath{\beta}}$.
 
We can solve equation (\ref{eq:pertEq}) for the potential correction, $\delta \psi(\bmath{\theta})$, provided that we start at a potential model that is close to the true potential. (We quantify what ``close'' means in the next section.)  One method to solve for the potential correction is to integrate along the characteristics of the partial differential equation (\ref{eq:pertEq}).  The solution is 
\be \label{eq:dpsi} 
\delta \psi (\bmath{\theta}) = \delta \psi(\bmath{\theta_{A}}) + \int^{\bmath{\theta}}_{\bmath{\theta_{A}}} \frac {d\theta_s \delta I (\bmath{\theta})}{\left\vert \frac{\partial I(\bmath{\beta})}{\partial \bmath{\beta}} \right\vert},
\ee
where
\be \label{eq:xs}
d\theta_s = \left( d\theta_1^2 + d\theta_2^2 \right)^{1/2},
\ee
\be \label{eq:dIdy}
\left\vert \frac{\partial I(\bmath{\beta})}{\partial \bmath{\beta}} \right\vert = \sqrt{\left( \frac{\partial I(\bmath{\beta})}{\partial \beta_1} \right)^2 + \left( \frac{\partial I(\bmath{\beta})}{\partial \beta_2} \right)^2},
\ee
and $\bmath{\theta_A}$ is an arbitrary reference point that is conveniently chosen to be at the location of one of the images, say A. (The reference point is arbitrary because the potential is determined up to a constant.)  The characteristic curves, on which we must integrate to obtain the potential correction, are given by curves that satisfy
\be \label{eq:charCurve}
\frac{d\theta_1}{d\theta_2} = \frac{\partial I / \partial \beta_1}{\partial I / \partial \beta_2}.
\ee
Each point on a characteristic curve thus follows the source intensity gradient (evaluated at the corresponding source location given by the lens equation (\ref{eq:lensEq})) that is directly translated to the image plane without distortions via the magnification matrix.  Due to the direct translation of the source intensity gradient, the characteristic curves differ from the curves on the image plane that map to the source intensity gradient curves.  The structure of the characteristic curves allows us to determine whether the potential solution given by equation (\ref{eq:dpsi}) is unique.  This is demonstrated in the next section.
 
We can repeat the process for a perturbative and iterative potential reconstruction method.  We expect the potential to be closer to the true potential after each iteration, which is indicated by a decrease in the magnitude of the intensity deficit.

The potential reconstruction method is non-parametric.  We can pixellate the potential distribution to match the observed image pixellation, and the potential correction at each pixel is given by equation (\ref{eq:dpsi}).  

To summarise, the four steps for the method are: (i) start with a potential model close to the true potential, (ii) calculate the intensity deficit (equation (\ref{eq:dI})) of each pixel, (iii) calculate the potential correction of each pixel (equation (\ref{eq:dpsi})) by integrating along the characteristics (equation (\ref{eq:charCurve})), (iv) obtain the corrected potential and repeat the process (steps (ii) to (iv)) until the intensity deficit approaches zero.  In the next section, we examine a quadruply imaged toy model to test the method of potential reconstruction.

\subsection{Example Toy Model}
To demonstrate the method of potential reconstruction discussed in the previous section, we consider a toy model with a simple lens potential that produces a quad like B1608+656.  

The toy system has a non-singular isothermal ellipsoid lens whose potential takes the form:
\be \label{eq:toyPot}
\psi(\theta_1,\theta_2) = (\theta_1^2 + 2 \theta_2^2 + 0.1)^{1/2}
\ee
We take the perturbed potential to be the original potential that is rotated clockwise by 1.1 degrees.  The source intensity distribution has elliptical contours with axis ratio of 0.634 and position angle of 147.2 degrees.  The source nucleus is located at $(\beta_{1s}, \beta_{2s})=(0.1, 0.05)$ and has an intensity peak of 100, in arbitrary units.  We assume the data is perfect with no noise, but we discretize the image plane region [-2,2]$\times$[-2,2] into a 201$\times$201 grid in order to correct for the perturbation of every pixel.  In Fig.~\ref{fig:toyConnChar}, the left panel shows the caustic curves (dashed) of the original potential and the source intensity contours (dotted), and the right panel shows the corresponding critical curves (dashed) and image intensity contours (dotted).  Analogous to B1608+656, there is an astroid caustic in the left panel.  The additional elliptical caustic curve is due to the non-singular nature of the lens potential.  Different regions separated by the caustic curves have different image multiplicities.  In the enclosed region intersected by the astroid and elliptical caustic curves, a source has five images on the image plane.  In the region within the caustic curves excluding the intersection, a source has three images.  In the region outside the caustic curves, a source has one image.  The astroid caustic is mapped to the outer critical curve and the elliptical caustic is mapped to the inner critical curve.  As for B1608+656, we focus on the astroid caustic and the outer critical curve.  Among the isophotes in the right panel, the four isophotal separatrices that are shown match to the four isophotes tangent to the astroid caustic in the left panel.  The separatrices intersect at the outer critical curve, as required (section 2.2.4).  Fig.~\ref{fig:toyTD} shows the arrival time delay contour of the source nucleus of the toy model.  The quad has similar time delay extrema (two saddles within the critical curve and two minima outside the critical curve) to the SPLE1+D(isotropic) model of B1608+656.

\begin{figure*}
\includegraphics[width=84mm]{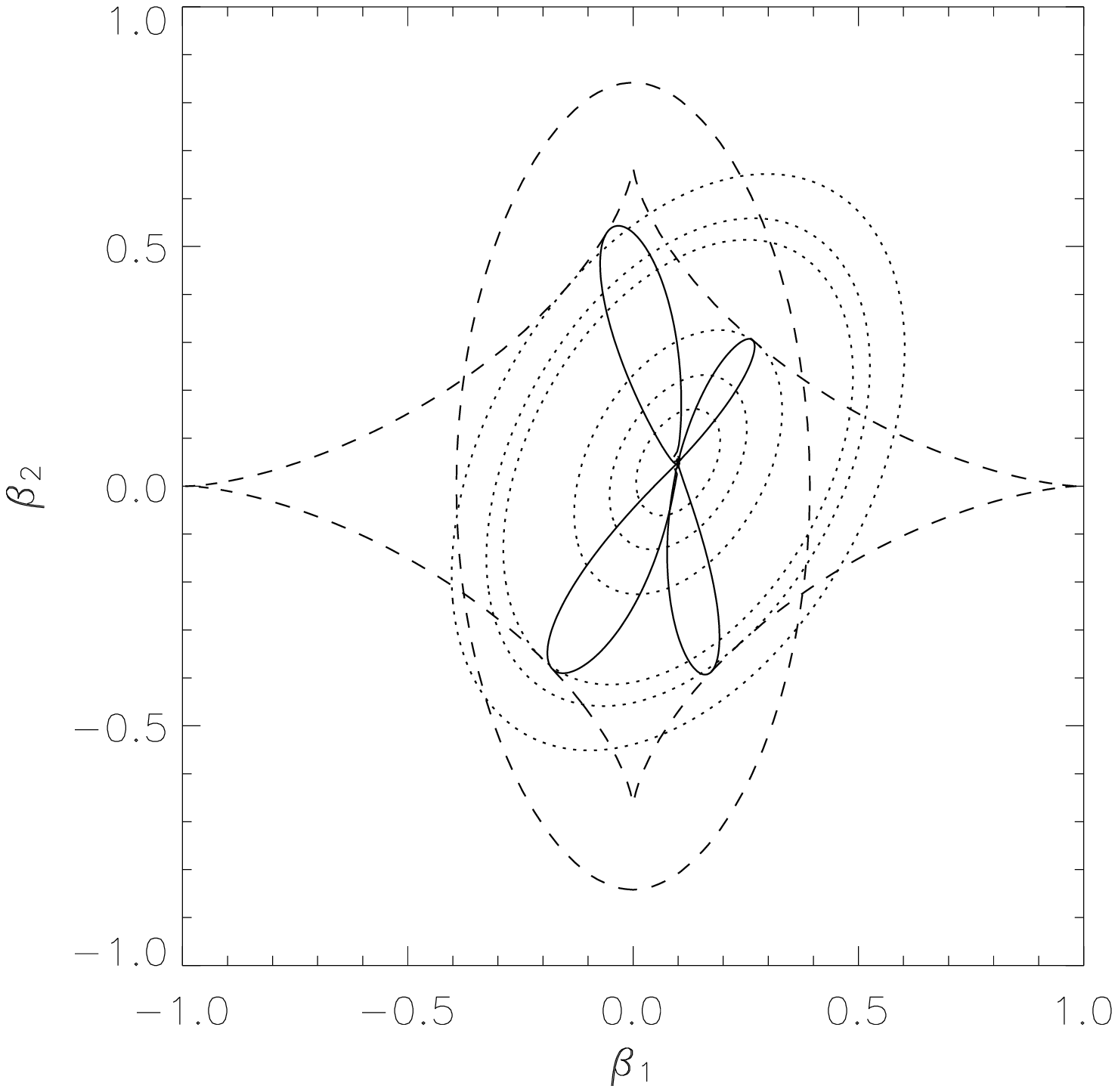}
\hspace{0.2in}
\includegraphics[width=84mm]{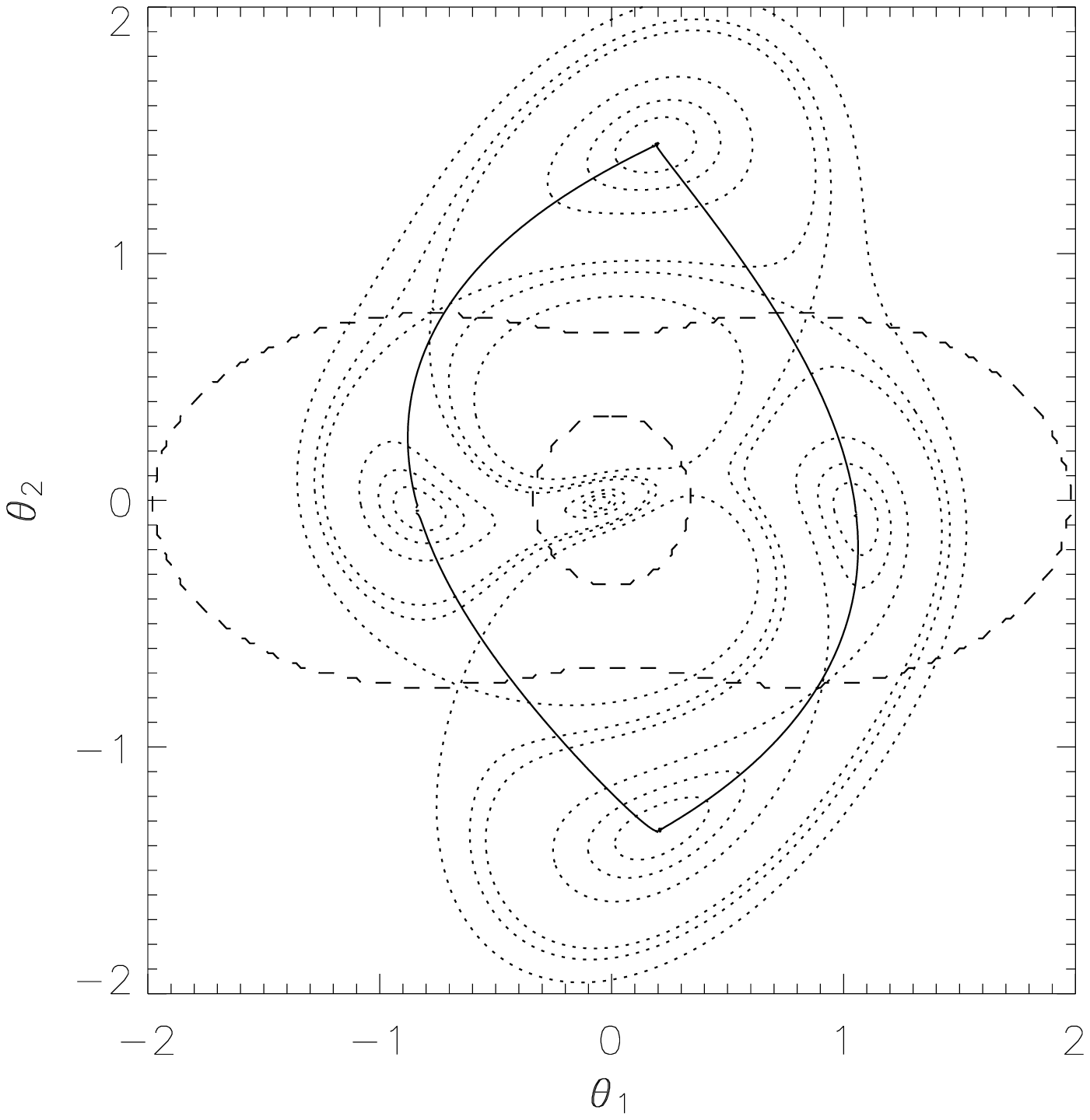}
\caption{\label{fig:toyConnChar} Left: the caustic curve (dashed) of the original toy potential model with the intensity contours (dotted) of the source.  Four of the intensity contours are tangent to the caustic curves.  The four mappings of the connecting characteristics (solid) are each tangent to the caustics.  Right: the critical curves (dashed) of the original toy potential model and the image intensity contours (dotted), four of which are isophotal separatrices intersecting at the outer critical curve.  The four connecting characteristics (solid) between the four images each cross the outer critical curve once.}
\end{figure*}

\begin{figure}
\includegraphics[width=84mm]{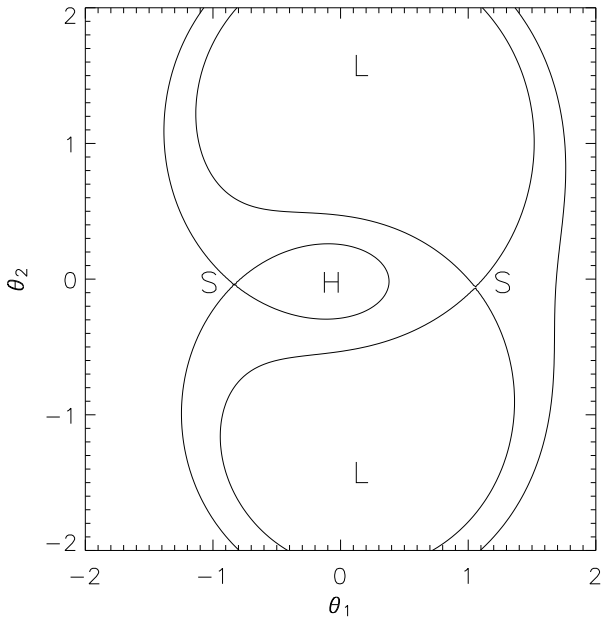}
\caption{\label{fig:toyTD} The time delay contour associated with the source nuclear position of the toy model.  The image locations of the source nucleus (see Fig.~\ref{fig:toyConnChar} right panel) are at time delay saddles (S), minima (L) or maxima (H).}
\end{figure}

We simplify the potential correction method by using the original source intensity distribution and the characteristic fields of the original potential (instead of reconstructing from the perturbed potential).  In reality, we would have to use the reconstructed source \citep{WD03} and the characteristic fields of the perturbed potential.  This would involve simultaneously determining the source and lens potential distributions and investigating the partial degeneracy between them, which are beyond the scope of this paper. We use the simplifying assumptions on the source intensity and characteristic curves as the first step to testing the method of potential reconstruction via integration along characteristics.  Only if the method works robustly in this simplified regime is the consideration of the more general problem relevant. 

Fig.~\ref{fig:toyChar} shows the characteristic field given by equation (\ref{eq:charCurve}).  The field has ``attractors'' (where field lines come together) and ``repellors'' (where field lines curve away) at the image locations of the source nucleus.  Using equation (\ref{eq:T}) and noting that the Jacobian matrix of $T(\bmath{\theta},\bmath{\beta})$ with respect to $\bmath{\theta}$ is equivalent to $\mathbfss{A}$ in equation (\ref{eq:jacob}) up to a constant coefficient, one can show that the attractors (or repellors) are associated to time delay minima/maxima (or saddles) for a source distribution that has non-crossing intensity contours.  A comparison between Fig.~\ref{fig:toyTD} and Fig.~\ref{fig:toyChar} confirms this fact.

\begin{figure}
\includegraphics[width=84mm]{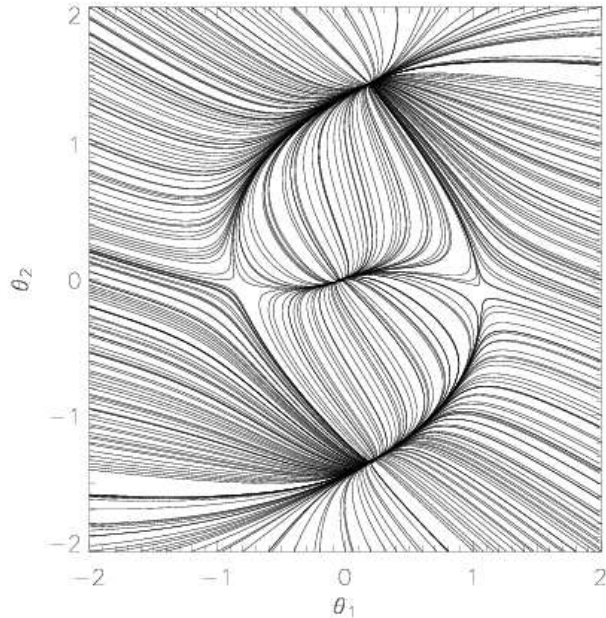}
\caption{\label{fig:toyChar} The characteristic fields of the toy potential model.  The attractors are associated with images that are time delay minima/maxima and the repellors are associated with time delay saddles.}
\end{figure}

We need to follow along the characteristics to correct for the potential perturbation given by equation (\ref{eq:dpsi}).  In Fig.~\ref{fig:toyChar}, almost all of the characteristic curves end at one of the three attractors; but there are special characteristic curves that connect the attractors and repellors.  These four \textit{connecting characteristics} between the four images (excluding the central image), shown in the right panel of Fig.~\ref{fig:toyConnChar} in solid lines, allow us to fix the potential offsets between the images and hence uniquely determine the potential up to a constant.  The left panel of Fig.~\ref{fig:toyConnChar} shows the mapping of these connecting characteristics onto the source plane (solid lines).  As one may expect, the mapping of each of the connecting characteristics between an attractor and a repellor is a loop on the source plane that is tangent to the astroid caustic curve due to the connecting characteristics intersecting the outer critical curve.

In addition to the characteristic curves, the intensity deficit is required for potential correction in equation (\ref{eq:dpsi}).  To get the intensity deficit defined in equation (\ref{eq:dI}) for the pixels on the image plane, first we use the perturbed potential model, the lens equation (\ref{eq:lensEq}), and the original source intensity distribution to get $I(\bmath{\beta})$, then we subtract it from $I(\bmath{\theta})$ obtained from the original potential.  Fig.~\ref{fig:dIPP012} shows the initial intensity deficit and the initial potential perturbation ($\delta \psi(\bmath{\theta})$ in equation (\ref{eq:potmodel})) before the perturbative and iterative potential correction, in the top left and bottom left panels, respectively.  We use plots of $\delta \psi(\bmath{\theta})$ to check that the perturbation approaches zero after corrections.

\begin{figure*}
\includegraphics[width=170mm]{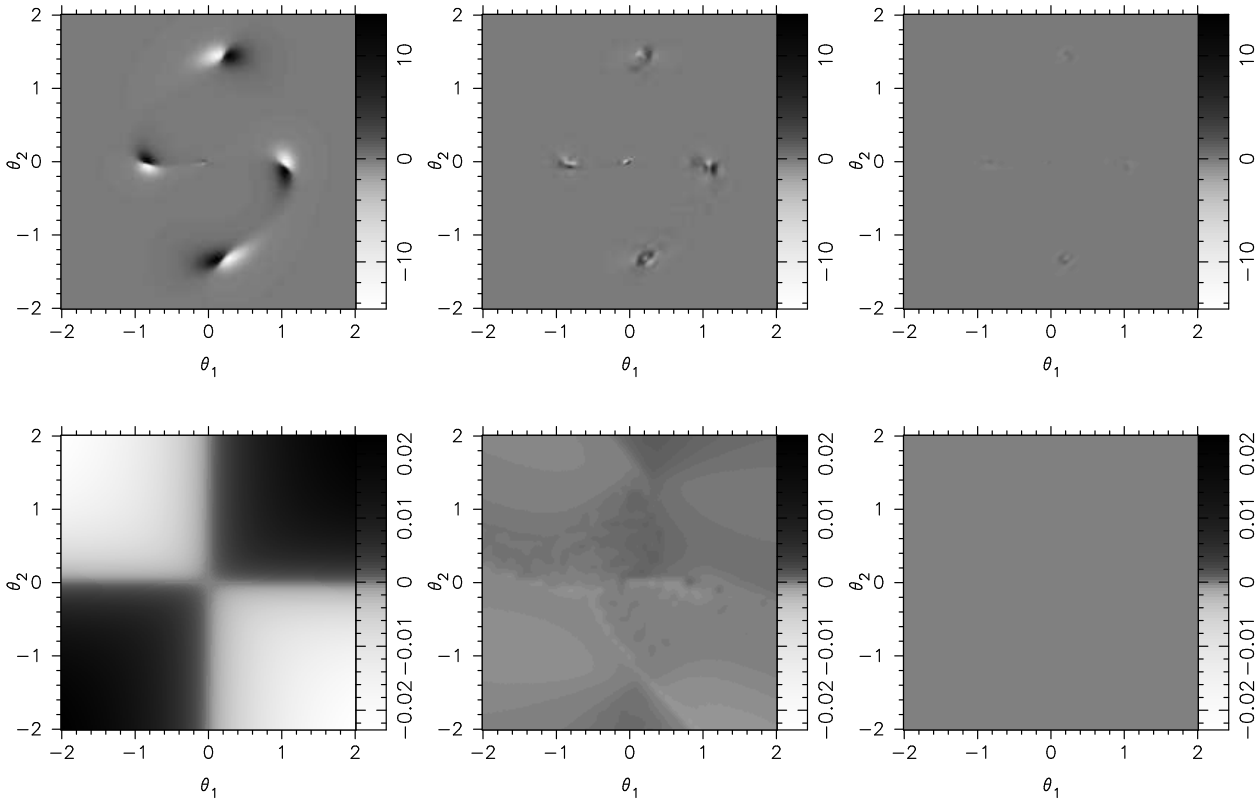}
\caption{\label{fig:dIPP012} Top row, left to right: the intensity deficit before potential correction, after 1 iteration and after 2 iterations of potential correction.  The maximum initial intensity deficit (top left) is 14 near the image positions (the peak nuclear source intensity is 100).  Bottom row, left to right: potential perturbation before correction, after 1 iteration and after 2 iterations of correction.  The initial potential perturbation magnitude (bottom left) is on average around 0.5 per cent of the original potential.  Since the potential is determined up to an arbitrary constant, the potential perturbation is plotted with respect to the mean to enhance small scale features.  The plotting scales of the middle and right panels (after corrections) are the same as the left panels (before corrections) for comparison.}
\end{figure*}

In each potential reconstruction iteration, we use the current perturbed potential model to obtain the intensity deficit ($\delta I(\bmath{\theta})$) and the source intensity gradient ($\left\vert \frac{\partial I(\bmath{\beta})}{\partial \bmath{\beta}} \right\vert$) at every pixel on the image plane; we then use equation (\ref{eq:dpsi}) to correct the perturbed potential by integrating along the characteristic curves of the original potential model.  Two iterations are performed and the resulting intensity deficit and potential perturbation after each iteration are shown in Fig.~\ref{fig:dIPP012}.  The middle and right panels show the intensity deficit (potential perturbation) in the top (bottom) after 1 and 2 iterations, respectively.  The middle and right panels are plotted on the same scales as that in the left panels.  Comparing the right panels to the left panels, the intensity deficit and potential perturbation converge to zero after two iterations (apart from numerical error), signifying that the method of potential reconstruction along characteristics works in theory with perfect data.

A possible limitation to this method is that the intensity deficit needs to be zero at the image locations; otherwise, according to equation (\ref{eq:dpsi}), the integrand diverges at the image locations, which are the end points of integration.  For the example above, we are saved from this divergence by discretizing the image plane and thus only reaching the image points within some tolerance, but never ending at the image (divergent) points.  The potential correction is most significant near the image points for any non-zero intensity deficit in the region.  Therefore, integrating along the characteristics may place limitations on the magnitude of potential perturbation that we can correct, which we discuss in the next section.

This method of potential reconstruction works only for small potential perturbations like the example we considered where the perturbation magnitude is on average (over the image grid) 0.5 per cent of the original potential.  By increasing the rotation of the original potential distribution to get the perturbed potential (that is, increasing the perturbation), we require more iterations for convergence, as expected.  When the rotation the original potential gets to $\sim 4.5$ degrees, which corresponds to an average potential perturbation magnitude of $\sim 1.5$ per cent, the method ceases to converge.  Therefore, the method of potential correction by integrating equation (\ref{eq:dpsi}) along characteristics works in theory with perfect data with a small ($\la$ 1 per cent) potential model error.  Unless a better algorithm is found for treating larger potential perturbation and real data with noise, the method proposed by \citet{BSK01} will not in practice be useful.

The example toy model considered provides a practical insight into the theory of potential reconstruction.  In reality, we do not have useful data everywhere due to the presence of noise; for an extended source, we can observe emission in an Einstein ring connecting the four images.  Based on the analysis of this section, the Einstein ring must be large enough to enclose the connecting characteristics in order to obtain proper potential offsets between the images.  This condition must hold for any potential reconstruction algorithm based on equation (15).

\section{Conclusions and further work}
We have considered the gravitational lens system B1608+656 to investigate the properties of quads.  We have defined limit curves as the loci of non-merging images as the source traces the caustic curve.  For the typical astroid caustic curve of a quad, there are inner and outer limits (relative to the critical curve) that are each tangent to the critical curve twice.  We have shown that isophotes that are tangent to the astroid caustic curve on the source plane map to isophotal separatrices on the image plane.  These separatrices must intersect on the critical curve and their associated satellite isophotes must be tangent to the limit curves.  We have shown that the current model proposed by \citet{K03} for B1608+656 does not satisfy these qualitative constraints for some of the isophotal separatrices.

We have investigated a perturbative and iterative method of potential reconstruction proposed by \citet{BSK01}.  The method requires solving a partial differential equation for the potential correction, which we have done by integrating along the characteristic curves.  We have used a toy model that is a quad like B1608+656 to test the method, assuming perfect data.  For small perturbations whose magnitudes are on average $\la$ 1 per cent of the original potential, the method has worked and we have had the perturbed potential converging to the true potential.  However, the method has failed to converge when the perturbation magnitude increases to around 1.5 per cent of the original potential.  This may be due to the non-zero intensity deficit near the image locations which restricts the integration along characteristics.  We hope to use the knowledge we have acquired about the anatomy of the quads and the characteristic fields of the potential correction equation to find a more robust method of potential correction that can be applied to real data with noise.

\section*{Acknowledgement}
We thank C. Fassnacht, L. Koopmans, P. Marshall, and T. Treu for useful discussions and encouragement.  We thank the referee for comments that improved the presentation of the paper.  SS thanks I. Mandel for commenting on the manuscript.  This work was supported by the NSF under award AST04-44059 and in part by the U.S. Department of Energy under contract number DE-AC02-76SF00515.  SS acknowledges the support of the NSERC (Canada) through the Postgraduate Scholarship.

\bsp

\label{lastpage}

\end{document}